\documentclass[10pt,aps,twocolumn,secnumarabic,amssymb, nobibnotes,  prx,superscriptaddress,floatfix]{revtex4-2}

\usepackage{graphicx}
\usepackage{dcolumn}

\usepackage{bm}
\usepackage[caption=false]{subfig}
\usepackage[colorlinks=true, citecolor=blue, linkcolor=blue, urlcolor=blue]{hyperref}
\usepackage{pinlabel} 
\usepackage{amsmath}
\usepackage{algorithm}
\usepackage{algpseudocode}
\usepackage{natbib}
\usepackage{braket}
\usepackage{physics}
\usepackage{color}
\usepackage[toc]{appendix} 

\usepackage{amssymb}
\usepackage{mathrsfs}
\usepackage[T1]{fontenc}

\usepackage{color}

\usepackage{float}

\begin{document}
\title{Information Transport in Classical-Quantum Hybrid System}

\author{Julian Rapp}
 \affiliation{Peter Gr\"unberg Institute,  Forschungszentrum J\"ulich, 52428 J\"ulich, Germany}
\author{Radhika H. Joshi}
 \affiliation{Peter Gr\"unberg Institute,  Forschungszentrum J\"ulich, 52428 J\"ulich, Germany}
\author{Alwin van Steensel}
 \affiliation{Peter Gr\"unberg Institute,  Forschungszentrum J\"ulich, 52428 J\"ulich, Germany}
\author{Yuli V. Nazarov}
 \affiliation{Kavli Institute of Nanoscience, Delft University of Technology, 2628 CJ Delft, The Netherlands}
\author{Mohammad H. Ansari}
 \affiliation{Peter Gr\"unberg Institute,  Forschungszentrum J\"ulich, 52428 J\"ulich, Germany}
 
\begin{abstract}
Many of the quantities that matter in quantum technology, e.g. entropy, entanglement, etc., cannot be associated with operator observables, because they depend non-linearly on the quantum state’s density matrix. The non-linearity provokes another bigger issue: Standard open-system equations evolve just one copy of that matrix, so we cannot track how such elusive quantities change. A recent formalism by Ansari and Nazarov sidestepped this by evolving multiple virtual replicas at once, but only in the weak-coupling regime. Here, we remove that limitation. Our generalized, multi-replicative master equation follows entropy flow and other vital metrics even when a quantum device interacts strongly with a classical environment. We show quantum coherence and hybridization jointly act to inhibit the net transfer of entropy, effectively introducing a ``bottleneck'' in the thermodynamic process. This sharper view of quantum–classical hybridization points the way toward more robust, resource-aware quantum hardware.  
\end{abstract}
\maketitle

\section{Introduction}

Entropy bridges quantum information science and quantum thermodynamics, serving both as a benchmark for quantum information resources and a limit on thermodynamic efficiency, a key measure for quantum technology such as quantum computers and ultra-efficient heat engines. \cite{BennettBernstein,ZulkowskiDeWeese,TouilDeffner,BallarinMangini}. In quantum heat engines, entropy governs operating principles and ultimately constrains efficiency~\cite{KoyanagiTanimura,HerreraReina,MoreiraSamuelsson,MatosdeAssis}. A precise understanding of how entropy is \emph{produced} is critical to unifying the thermodynamic and information-theoretic descriptions of quantum devices~\cite{SantosCeleri,DolatkhahSalimi,Salazar,xu2023parasitic-free,PtaszynskiEsposito,LacerdaKewming,WebbStafford,khoshnegar2014toward}. Real-time entropy measurement cannot take place by physical observables--such as energy, charge, and spin--because of the nonlinearity of the density matrix in the definition of entropy, makes it a non-observable quantity. However, entropy production can be tracked in the statistics of physical currents \cite{KlichLevitov,AnsariNazarovCorrespondence}. 
 
Although quantum devices exploit superposition and entanglement--phenomena absent from classical stochastic systems--early studies found no intrinsically quantum signature in entropy flow beyond the indirect influence of coherence on transition probabilities. In these models, the entropy current is governed solely by excitation and decay probabilities (see \cite{arXiv2018AlickiKosloffIntroductiontoQuantumThermodynamicsHistoryandProspects, 101103physrevx50310442015PhysicalReviewXUzdinKosloffEquivalenceofQuantumHeatMachinesandQuantumThermodynamicSignatures}), effectively reducing quantum dynamics to a classical stochastic description. 

The pathology is in the shortcut of plugging the quantum-master-equation solution into the definition of entropy; that equation evaluates the time evolution of a single copy of density matrix, while entropy requires a time differential equation written for more than one copy of density matrix; otherwise, the shortcut hides coherence contributions and systematically underestimates/overestimates entropy production.  

A significant advance in the theoretical understanding of entropy flows was achieved by introducing a Keldysh-based diagrammatic  formalism~\cite{keldysh1965similar,AnsariNazarov16,Nazarov,AnsariNazarovEngines}. This formalism, namely Keldysh-Ansari-Nazarov (KAN),  uniquely generalizes the traditional Keldysh contour approach by replicating contours, analogous to the replica trick commonly employed in quantum field theory~\cite{HolzheyLarsen,CalabreseCardy}. This embeds the intrinsic nonlinearity of entropy into a comprehensive diagrammatic framework constructed upon nonequilibrium Green functions. The new formalism is a consistent generalization of the quantum master equation to quantities that are non-linear in the density matrix \cite{UtsumiFluctuation,UtsumiPRB,Reimer2019}. This has unveiled a novel quantum effect, the quantum-coherence-mediated entropy transfer \cite{UtsumiPRB}, which challenges classical and stochastic results, emphasizing the essential role of coherence in the thermodynamics of a quantum system, which is missing from earlier literature. 
\begin{figure}[h!]
    \includegraphics[width=0.98\columnwidth]{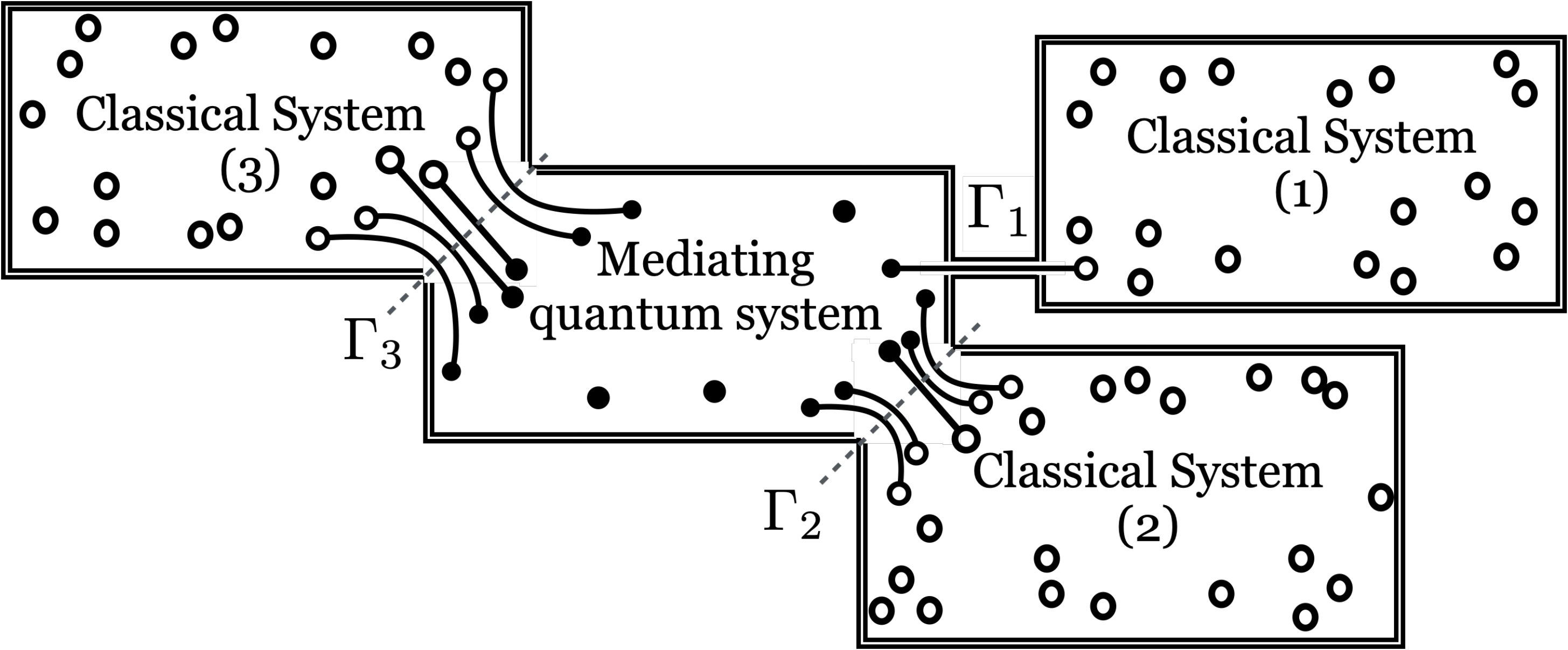}
    \vspace{-0.2cm}
\caption{A quantum system mediating the flow of information between three classical systems hybridized to systems (2) and (3) and  weak coupled to (1). }
\vspace{-0.2cm}
\label{fig.diag}
\end{figure}

Subsequent developments established how to measure entropy by R/FCS correspondence--an exact correspondence between entropy flow and the full statistics of energy transfer ~\cite{Ansari,AnsariSteenselNazarov,AnsariNazarovCorrespondence}. This discovery demonstrated how quantum superposition can impact the statistics of energy transfer corresponding to the coherent entropy flow. The emergence of a superposition-dependent entropy flow and its direct connection to the complex quantum statistics of energy transfer has the potential to heavily affect a significant advancement for quantum circuit theory~\cite{Circui,pazem2023err,xu2024lattice, ku2020suppression,xuzzfree}, by providing crucial insights for accurately modeling quantum nanostructures \cite{blais2020circuit, koch2007charge-insensitive, Kindermann,ansari2013the-effect,bal2015dynamics} and controlling the flow of information and physical quantities in them \cite{martinis2020quantum,devoret2013superconducting,xu2024lattice,xu2025,pazem2023error,ansari2011noise}.
 
In this study we present a unified framework that treats system–environment coupling at any strength—an essential step for solid-state platforms, where quantum–classical hybridization is the rule rather than the exception. By casting the multi-replica master equation into a generalized tensor-network language, we can follow R\'enyi–entropy currents without assuming that subsystems remain strictly separable. Figure \ref{fig.diag} sketches the scenario: a single quantum node channels information to three classical subsystems. Subsystem (1) interacts weakly via a small transition rate $\Gamma_1$, so the original KAN formalism suffices. Subsystems (2) and (3), however, share an extensive set of degrees of freedom with the quantum node, defying any diagonalization and perturbation treatment and demanding our new hybridization extension to KAN formalism.

The paper is organized as follows. In Sec. II, we revisit weak-coupling theory, illustrating it with two examples in which a qubit and a harmonic oscillator act as energy valves. Section III introduces the concept of quantum–classical hybridization, presenting the time evolution of a generalized density matrix together with its diagrammatic Keldysh-contour formulation. In Sec. IV, we analyze a hybridized qubit as an energy valve, first in a two-replica setting, then extending the framework to M replicas, and finally summarizing the results for the von Neumann entropy in the hybrid system. We conclude by noting that our approach bridges the methodological gap between weak and strong coupling, enabling accurate entropy quantification in realistically hybridized quantum devices.

\section{Weak Coupling Formalism}
\label{S: Old Formalism}


The canonical quantum measure of entropy is the von Neumann entropy
\begin{equation}\label{eq:vN}
  S_{\mathrm{vN}}
  =-k_{\mathrm{B}}\operatorname{Tr}\bigl(\rho\ln\rho\bigr),
\end{equation}
with \(k_{\mathrm{B}}\) the Boltzmann constant.  
Because \(S_{\mathrm{vN}}\) is a nonlinear functional of the density matrix \(\rho\), its instantaneous flow cannot be accessed by a single physical observable; instead it is inferred indirectly from measurable energy exchanges~\cite{KlichLevitov}.

\paragraph*{R\'enyi--entropy.}---For many practical tasks it is advantageous to work with the R\'enyi entropy of integer rank~\(M\), a natural extension of the von Neumann entropy that is conserved in closed systems and widely used to quantify multi-qubit entanglement \cite{Rastegin,HuangYin,LopezKos,TarabungaTirrito}, probe nonequilibrium dynamics \cite{PainRoy,BulgarelliPanero,CalabreseCardy,SohalNie,RuggieroTurkeshi}, and characterise quantum many-body states \cite{science.Esslinger.Monitoring.Goldstone}. In quantum theory one replaces classical probabilities with the density matrix~\(\rho\) and defines the R\'enyi entropy as \(S_M(\rho)=\tfrac{k_{\mathrm{B}}}{1-M}\ln[\operatorname{Tr}(\rho^{M})]\) with integer \(M>1\). The factor \(1-M\) ensures that the von Neumann entropy \(S_{\mathrm{vN}}=-k_{\mathrm{B}}\operatorname{Tr}(\rho\ln\rho)\) is recovered smoothly in the replica limit \(M\to1\): both numerator and denominator vanish, and applying L'H\^opital’s rule to the logarithm yields the familiar expression.  For practical calculations, Ref.~\cite{Nazarov} adopts a simplified and equivalent definition,
\begin{equation}
S_M=k_{\mathrm{B}}\,\mathrm{Tr}\bigl(\rho^M\bigr),
\label{eq:Renyidef}
\end{equation}
which retains the essential polynomial structure while sidestepping the $\ln$ and the explicit $1-M$ factor. In this work we follow that convention. 

\paragraph*{R\`enyi--entropy flow.}---The associated entropy flow is defined as
\begin{equation}
F_M=\frac{1}{S_M}\,\frac{dS_M}{dt}, \qquad F_{\mathrm{vN}}=\lim_{M\to1}\frac{dF_M}{dM}.
\label{eq.Renyiflow}
\end{equation}
In a \emph{closed} system the R\'enyi entropy is conserved. Observing only a \emph{part} of a composite system converts this global conservation law into a current of R\'enyi entropy that flows between subsystems~\cite{Nazarov}. We study a quantum system that mediates information between several reservoirs. A reservoir acting as a classical probe, labeled by the index \(b\), is a classical environment held at a fixed, low temperature and is coupled to the system through a small transition rate \(\Gamma_{b}\) that weakly perturbs the quantum system \cite{ansari2015rate}.  For the purpose of evaluating the entropy exchanged with the probe, we work in the weak--coupling regime \(\Gamma_{b}\ll\Gamma_{e}\); here \(\Gamma_{e}\) collectively denotes the couplings to all other reservoirs \(e\).  All reservoirs are maintained in the linear--response regime: they remain at fixed temperature, do not interact directly with each other, and any nonequilibrium effects arise solely within the quantum node as a shared small subsystem between them.  The reduced dynamics under the couplings \(\Gamma_{e}\) are obtained by solving an appropriate open quantum system approach and tracing out every environment.  The resulting density matrix governs the time evolution of the quantum node \(\rho(t)\), which is then substituted into Eq.~\eqref{eq.Renyiflow} to compute the R 'enyi entropy current flowing into the probe reservoir \(b\).

Since the total R\'enyi entropy of the composite system is conserved, the resulting entropy current depends only on the instantaneous state of the central subsystem, which therefore acts as an entropy ``valve.''  Evaluating the flow of entropy from Eq. (\ref{eq.Renyiflow}) requires a replicated quantum master equation. More precisely,  \(M\) replicas of the closed system are considered: for both the quantum system and the probe reservoir, separate closed-timeline Keldysh-type contours are considered in each replica with backward (forward) flow for bra (ket) states. Since the measurement takes place in the probe reservoir, the contour associated with the quantum system gets closed in each replica to denote the tracing of the quantum system out, leaving only the probe reservoir branches connected across replicas. This results in the quantum master equation for a multi-replicated density matrix of probe reservoir, which after proper final tracing, will result in the entropy flow  to probe reservoir. 

Detailed analysis in \cite{AnsariNazarovEngines} shows two parts of entropy flow emerge: a previously-known incoherent part due to stochastic level transitions, and a direct coherent part due to quantum superpositions:
\begin{equation}
  F_M = F_M^{\mathrm{incoh}} - F_M^{\mathrm{coh}}.
  \label{eq. total weak entropy}
\end{equation}

In the absence of stationary quantum coherence, the expression are reduced to standard energy-exchange statistics.   Persistent coherence can make the net flow negative, a clear signature of genuinely quantum back-action among the replicas.

{\bf Example 1}:  Consider a simple quantum system consisting of a qubit with two states, $|0\rangle $ and $|1\rangle$, separated by energy $\hbar\omega=(E_1 - E_0)$. This qubit is driven by an oscillating external field described by the Hamiltonian  $H_{dr}=\hbar \sum_{m,n=\{0,1\}}\Omega |m\rangle \langle n | \exp(-i \omega_{dr} t) +H.c.$, with $m \neq n$ and the driving frequency $\omega_{\text{dr}}$ close to the qubit's resonance frequency $\omega$ and amplitude $\hbar\Omega$. The qubit mediates energy and information exchange between two reservoirs: a weakly coupled probe reservoir at nearly zero temperature and a stronger-coupled environment at finite temperature $T_b$.

In the linear response regime, each reservoir interacts with the qubit via the coupling Hamiltonian $H_{\text{int}}=\hbar \sum_{m,n=0,1}|m\rangle \langle n| \hat{X}_{mn}^{\alpha}$, where $ \hat{X}_{mn}^{\alpha}$ acts within the Hilbert space of reservoir $\alpha=\{b,e\}$. For an energy exchange quantum of $\Omega_P \equiv P \omega$, the excitation (absorption) rate, derived using the Kubo-Martin-Schwinger (KMS) relation \cite{KMS,AnsariNazarovEngines}, is $\Gamma_{\uparrow}^{\alpha}(\Omega_P/T^{\alpha})= n_{\text{B}}(\Omega_P/T^{\alpha}) \chi^{\alpha}$, while the emission rate satisfies $\Gamma_{\downarrow}^{\alpha}(\Omega_P/T^{\alpha}) / \Gamma_{\uparrow}^{\alpha}(\Omega_P/T^{\alpha})= \exp(\hbar\Omega_P/k_{\text{B}}T^{\alpha})$. Here $n_{\text{B}}(\Omega_P/T^{\alpha})=1/(\exp(\hbar \Omega_P/k_{\text{B}}T^{\alpha})-1)$ is the Bose distribution and $\chi^{\alpha}\propto \Gamma_{\alpha}$ denotes the temperature-independent dynamical susceptibility. For simplicity, when considering only the fundamental energy quantum ($P=1$), we omit the explicit dependence on frequency and temperature.

Ref. \cite{AnsariNazarovEngines} shows that the stationary-state R\'enyi entropy flow into the probe reservoir comprises two distinct parts: the incoherent (probabilistic) contribution $F_M^{\text{incoh}}= G_M \omega(\Gamma^b_{\downarrow} p_1 - \Gamma^b_{\uparrow} p_0)$, and the coherent contribution $F_M^{\text{coh}}= G_M \omega \Gamma^b |\rho_{01}|^2$, with $p_{0,1}$ representing the populations and $\rho_{01}$ the coherence of the qubit density matrix in the energy basis; $\Gamma^b \equiv \Gamma^b_{\downarrow}- \Gamma^b_{\uparrow}$. The prefactor $G_M$ is defined as $G_M \equiv M n_{\text{B}}(\Omega_M / T)/ [n_{\text{B}}(\Omega_{M-1} / T) n_{\text{B}}(\Omega_1/T)]$. In the limit of $M\to 1$ one  recovers the von Neumann entropy for this system which turns out to still have the coherent flow persistent in it: 
\begin{equation}
    F_{\rm{vN}}= (\Gamma^b_{\downarrow} p_1 - \Gamma^b_{\uparrow} p_0 -  \Gamma^b |\rho_{01}|^2 ) \omega /T_{\rm{b}}
    \label{eq.qubitfvn}
\end{equation}

In the presence of stationary quantum coherence in the system, the toral entropy flow slows down but cannot become negative, which indicates that the probe is entropy sink, yet the qubit coherence slows down the exchange rate.

{\bf Example 2}: Consider a harmonic oscillator with frequency $\omega$, described by the Hamiltonian $\hat{H}=\hbar \omega (\hat{a}^\dagger \hat{a} + 1/2)$, externally driven at frequency $\omega_{\text{dr}}=\omega$. This oscillator mediates energy exchange between an environment at temperature $T_e$ and a weakly coupled probe reservoir at temperature $T_{\text{b}}$. Due to energy conservation, exchanges occur either at the oscillator frequency $\omega$, reflecting the oscillator's ability to respond resonantly both at its natural and external driving frequencies. Let $T_{\text{h}}$ denote the oscillator's effective temperature defined through correlations $\langle \hat{a}\hat{a}^\dagger \rangle = \bar{n}(\omega/T_{\text{h}})+1$ and $\langle \hat{a}^\dagger \hat{a}\rangle = \bar{n}(\omega/T_{\text{h}})$. Ref.~\cite{AnsariNazarovCorrespondence} demonstrates that the stationary-state R\'enyi entropy flow into the probe splits into two distinct parts as $F_M=G_M \chi^b [n_{\text{B}}(\omega/T_{\text{h}})-n_{\text{B}}(\omega/T_b)]$, with the prefactor $G_M$ defined previously. The corresponding von Neumann entropy flow simplifies to 

\begin{equation*}
    F_{\text{vN}}=[n_{\text{B}}(\omega/T_{\text{h}})-n_{\text{B}}(\omega/T_b)] \chi^b \omega/T_{\text{b}}
\end{equation*}
 
 Notably, the entropy flow depends solely on the oscillator and reservoir temperatures and remains unaffected by the external drive amplitude. The flow reverses sign at $T_b\geq T_{\text{h}}$, becoming negative when the quantum oscillator coherently injects entropy into the colder reservoir.

The examples discussed here and others such as photovoltaic cells \cite{Ansari, AnsariSteenselNazarov} illustrate that a consistent extension of information measures to quantum heat engines is feasible and can serve for pathfinding next generation of quantum heat engines. 

This paper aims to further extend this consistent theory into the regime of strong coupling, where significant probe-system interaction profoundly influences the system's quantum dynamics.

\section{Quantum-Classical Hybridization }

In a composite system consisting of a number of classical system coupled to a mediating quantum system, the textbook Hamiltonian is split as $H = H_{Q} + \sum_C H_{C} + \sum_C H_{\mathrm{QC}}$, with Quantum-Classical interaction \(\lVert H_{\mathrm{QC}}\rVert\) much smaller than the minimum energy gaps in the quantum Hamiltonian \(H_{Q}\) and the classical systems \(H_{C}\). One diagonalizes \(H_{Q}\) and \(H_{C}\) separately and treats \(H_{\mathrm{QC}}\) perturbatively.   Each reservoir, assumed macroscopic with a continuous spectrum, sits at its fixed temperature, introducing only small corrections to the isolated dynamics.
 \paragraph*{Breakdown of separability in the hybridized regime: }---We consider a device in which the quantum subsystem is \emph{strongly} correlated with its classical probing environment.  Here we consider the example of a low temperature classical system that serves as a {probe} measurement with transition rate $\Gamma_b$, and a second classical environment at finite temperature with transition rate $\Gamma_e$.

Strong coupling merges classical and quantum dynamics, so that the usual system-bath separation no longer applies.  Accurate descriptions must therefore include the hybridized degrees of freedom explicitly and compute transport and feedback. In Fig. (\ref{fig.diag}) the classical systems (2) and (3) are strongly coupled so that  treating them as distinct subsystems is inaccurate. In other words, estimating the subsystem energy spectrum in the absence and in the presence of other strongly-interacting subsystems will be completely different so that one will not be an approximation of the other one. What we present below is going to emphasize on the this distinction by comparing weak and strong coupling strengths. 

In the strong coupling case, the transition rates induced by every reservoir are comparable,
\begin{equation}
    \Gamma_{b}\;\approx\;\Gamma_{e},
\label{eq.strongrate}
\end{equation}
which indicates that the classical degrees of freedom intertwine with the qubit states and exert maximal control over the dynamics.

\paragraph*{Basis Construction for Hybridization.}---The R\'enyi entropy in Eq.~(\ref{eq:Renyidef}) \emph{implicitly assumes} that the $M$ replica density matrices are separable, sharing only the minimal degrees of freedom.  Whether among atomic orbitals in chemistry~\cite{PRXQuantum.6.020319}, phonon–photon polaritons in solids~\cite{PhysRevLett.121.227401}, or spin–orbit coupling in atomic systems~\cite{PhysRevLett.117.210402},, hybridization is the coherent mixing of basis states that were once subsystem‐specific.  Therefore, we work in a tensor product basis that spans \emph{entire} Hilbert space. The Hamiltonian takes a block‐structured (often sparse): diagonal blocks hold the isolated energies, while off-diagonal blocks encode the couplings that drive hybridization.  If these couplings dominate, the ensuing entanglement is so widespread that states can no longer be unambiguously assigned to any single component.

When three quantum subsystems C, Q, and $\mathrm{\overline{C}}$ are hybridized, their joint state resides in the composite Hilbert
space and is fully characterized by a density
operator $\rho\in\mathrm{L}(\mathcal{H}_C \otimes \mathcal{H}_Q \otimes \mathcal{H}_{\overline{C}})$, with $\mathrm{L}$ being the linear map within the Hilbert space to the density matrix representation.

\paragraph*{R\'enyi entropy of hybridized system.}---Combining the concept of tensor network time evolution can help redefine the entropy of R\'enyi for hybridized systems. The concept of separability of replicas in the original definition of R\`neyi entropy can be generalized, yet produces scalar. As a reminder, let us consider a probe reservoir $\mathrm{\overline{C}}$, a quantum system Q modulated by environment C. KAN formalism proposes that the probe entropy is determined from $\Tr_{\overline{C}} (\rho_{\overline{C}})^M$, which suggests that all contours devoting to Q and C should be traced out within each replica, except the probe density matrix, which encompasses all replicas and is then traced.

Between subsystems  weak coupling can set a boundary; however, hybridization subsystems makes the boundary dynamically ill-defined. To address this breakdown, consider the density matrix subject to time evolution of the quantum system and a partial subset of coupled classical system, e.g. $d\rho_{_{Q{\overline{C}}}}/dt=\mathcal{L}\rho_{_{Q{\overline{C}}}}$ with $\mathcal{L}$ being the so-called Liuovillian operator, a typically non-Hermitian operator to be determined perturbatively or using renormalization group methods.  

Generalizing the R\'enyi formalism to include hybridization calls for
all degrees of freedom that intertwine the subsystems, prompting the use
of a density matrix spanning \emph{all} $M$ replicas.  Instead of isolating $\rho_{\overline{C}}$, we consider entangled subsystems $\rho_{CQ\overline{C}}$, and instead of the simple product we
consider their tensor product. 
\begin{equation}
\rho_{(CQ\overline{C})_1(CQ\overline{C})_2\ldots(CQ\overline{C})_M} \equiv {\bigotimes_{i=1}^M}\rho_{({CQ\overline{C})_{i}}}
\label{eq. HRenyidef}
\end{equation}
with the index pattern repeated $M$ times.  For clarity, we denote the $i$-th replica of
$CQ\overline{C}$ in the tensor by $(CQ\overline{C})_{i}$. The symbol on the right side of Eq. (\ref{eq. HRenyidef}) is used for brevity of notation.

\begin{figure}[t]
    \includegraphics[width=1\columnwidth]{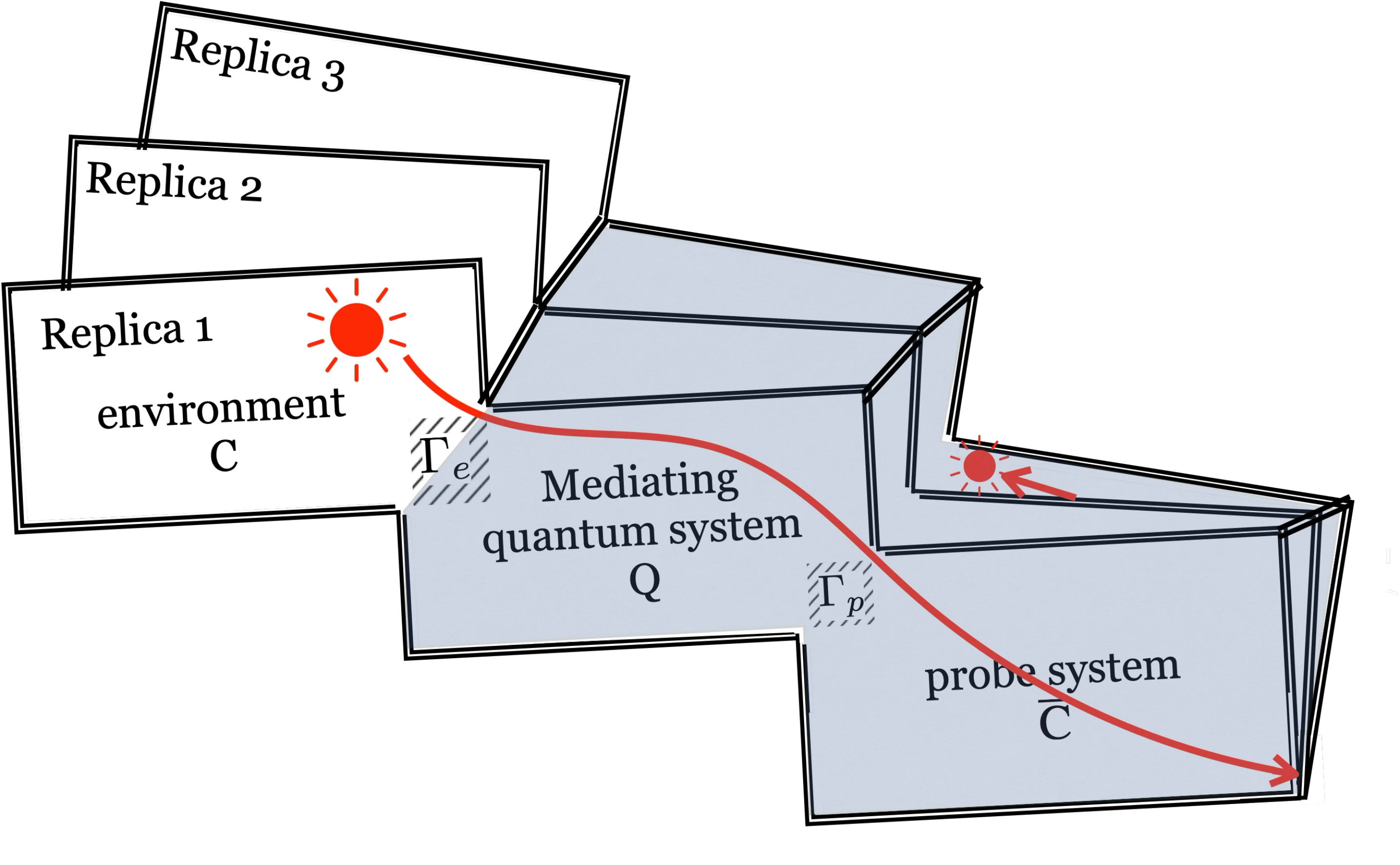}
\caption{Hybrid quantum bridge between reservoirs at temperatures $T_{C}$ and $T_{\overline{C}}$.  Three replicas are shown, and only the information flux into the probe reservoir ($\overline{C}$) is monitored.  The red path illustrates a cross-replica event: a particle enters from $C$, traverses the quantum mediator, reaches the probe, and is re-emitted into a different replica.}
\label{fig.repdiag}
\end{figure}

By following the real environment $C$ in each replica and tracing it out, the mutual density matrix is determined for hybridized $M$ replicas of the quantum system and the probe reservoir.

Let us now define the generalized replicated density matrix for  $M$
hybridized quantum systems only, after tracing out environment in each replia, the probe reservoir and the quantum system go together in all $M$ replicas. This allows not only exchanges of particles in one replica to take place, but also exchanges in between replicas becomes a possibility, similar to weak coupling regime. By tracing out the probe reservoir  we are left with $M$ replicas of the quantum system density matrix, which we denote as  
\begin{equation}
\label{E:Rhat_definition}
\hat{R}_{M}^{(\overline{C})}= \Tr_{\overline{C}}
\Bigl[\bigotimes_{i=1}^{M}\Tr_{C}\left(\rho_{(CQ\overline{C})_{i}}\right)\Bigr],
\end{equation}

$\hat{R}_{M}^{(\overline{C})}$ denotes the tensor product of $M$ replicas of the quantum system, where the superscript $(\overline{C})$ reminds us that the probe reservoir has already been traced out once, after experiencing the cumulative influence of all $M$ replicas. Performing a subsequent trace over the system’s degrees of freedom finally yields the following scalar quantity:
\begin{equation}
\label{E:Renyi_generalization}
S_{M}= \Tr_{Q}\Tr_{\overline{C}}
\Bigl[\bigotimes_{i=1}^{M}\Tr_{C}\!\bigl(\rho_{CQ\overline{C}_{i}}\bigr)\Bigr].
\end{equation}
which defines R\'enyi entropy in strong coupling regime for convoluted systems so that they cannot be identified as isolated system anymore.

The \emph{replicated composite system} is regarded as closed; hence transitions
can occur not only between the quantum system and the two classical reservoirs
within each copy, but also \emph{between} different replicas.  This structure
is illustrated in Fig.~\ref{fig.repdiag}, where the darker regions indicate the
channels through which replicas can exchange particles.

\subsection{Evolution of Generalized Density Matrix}
\label{sec. generalized dm}

Computing the R\'enyi entropy flow $dS_M/dt$ in strongly coupled systems from Eq. (\ref{E:Renyi_generalization}) requires a novel extension of the KAN diagrams in order to represent $d\hat{R}_M/dt$, i.e. the time evolution of the generalized density matrix defined  Eq. (\ref{E:Rhat_definition}).

We treat the evolution of $\hat{R}_M$ in Liouville space \cite{Gyamfi}, where we can write it in terms of the Liouvillian superoperator $\mathcal{L}_M$, which means it acts the matrix of generalized density matrix---rather than on state vectors, which is
\begin{equation}
\label{E: Liouville equation}
    \frac{\partial}{\partial t} \ket{R_M}\!\rangle = \mathcal{L}_M \ket{R_M} \!\rangle
\end{equation}
with $\ket{R_M}\!\rangle$ being the vectorized version of generalized density matrix defined in Eq. (\ref{E:Rhat_definition}).  The Liouvillian operator encodes \emph{both} coherent oscillations and
decoherence rates.  When diagonalized in its eigenbasis, its spectrum
generically acquires negative real parts that set those decay rates.  Let
$\{\ket{\Phi_0}\!\rangle,\ket{\Phi_1}\!\rangle,\ket{\Phi_2}\!\rangle,\ldots\}$ be an
operator basis composed of Liouvillian eigenvectors with corresponding
eigenvalues
\(\{\Lambda_{0},\Lambda_{1},\Lambda_{2},\ldots\}\):
\begin{equation}
\label{E: Eigendecomposition}
\mathcal L_{M}\,\ket{\Phi_i}\!\rangle = - \Lambda_{i}\,\ket{\Phi_i}\!\rangle.
\end{equation}

\noindent
The formal solution of the evolution equation~(\ref{E: Liouville equation})
can then be written as
\begin{equation}
\label{E: Eigenvalue evolution}
\ket{R_{M}(t)}\!\rangle  = \sum_{i} c_{i}\,e^{-\Lambda_{i}t} \ket{\Phi_i}\!\rangle .
\end{equation}

Because \(\mathcal L_{M}\) contains non-Hermitian dissipative parts, its
eigenvalues are in general \emph{complex}; the Liouvillian structure ensures
that they occur in complex-conjugate pairs.  Moreover, all real parts must be
non-positive: $\ket{R_{M}(t)}\!\rangle$ is expected to relax to a stationary form
independent of initial conditions, and any positive real part would cause an
unphysical divergence.

At long times all contributions with strictly negative real parts decay
exponentially, leaving only the term with the \emph{smallest} real part magnitude,
\(\Lambda_{0}\), \footnote{%
In principle \(\Lambda_{0}\) could be a complex pair.  For
\(\Gamma_{A}\ll\Gamma_{B}\) the system relaxes to a stationary state, implying
\(\Lambda_{0}=0\).  As \(\Gamma_{A}\) increases continuously,
\(\Lambda_{0}\) decreases smoothly and remains real; in none of our cases did
the leading eigenvalue become complex, which would signal a discontinuity.},
so that \(\hat R_{M}(t)\) can be approximated by \(|\Phi_{0}\rangle\!\rangle\) alone.

Transforming this late-time form of the generalized density matrix back to
state space and inserting the \(M\)-fold product into the R\'enyi definition
yields the entropy flow into the probe reservoir \(\overline C\):
\begin{align}
\label{E: Renyi conjecture}
F_{M}
 &= \frac{1}{S_{M}^{(\overline C)}}\,\frac{\partial S_{M}^{(\overline C)}}{\partial t}
    = -\,\frac{1}{\operatorname{Tr}_{Q}\hat R_{M}(t)}
        \operatorname{Tr}_{Q}\!\Bigl[\tfrac{\partial\hat R_{M}}{\partial t}\Bigr]
    = -\,\Lambda_{0}.
\end{align}
Hence the production rate of R\'enyi entropy depends only on the \emph{lowest}
(non-positive) eigenvalue of the Liouvillian that governs the time evolution of
the generalised \(M\)-replica density matrix~\(\hat R_{M}\).  The
trace is taken over the quantum subsystem acting on the corresponding
eigen-operator \(\Phi_{0}\).

\subsection{Diagrammatic Keldysh Contours}

Motivated by the definition of the R\'enyi entropy through the generalised density matrices in Eqs.~(\ref{E:Rhat_definition}) and (\ref{E:Renyi_generalization}), we can extend the KAN formalism from the weak- to the strong-coupling regime.  For this, we note that the final result in Eq.~(\ref{E: Renyi conjecture}) shows that the R\'enyi-entropy flow does \emph{not} require tracing out all contours; instead it is fixed by the \emph{lowest non-zero} eigenvalue of the Liouvillian operator.  

For a classical–quantum–classical hybridised system, illustrated in Fig.~\ref{fig.repdiag}, we consider \emph{three} contours in each replica.  Time runs from the distant past at the left to the present on the right.  Every contour first moves \textit{backwards} in time, representing a bra state $\langle\cdot|$, passes through the initial density matrix (depicted as a box with label $\rho_{-\infty}$), and then returns \textit{forwards} in time toward the present to represent the corresponding ket state $|\cdot\rangle$ of that subsystem.  

As indicated by Eq.~(\ref{E:Rhat_definition}) the trace over subsystem $C$ is performed \emph{within} each replica; this intra-replica trace is denoted by the green contour.  The trace over the probe subsystem $\overline{C}$, however, is taken \emph{after} wrapping around all replicas.  We depict this by an outer black contour, left open-ended (black arrowhead) to signal that it closes only after enveloping every replica; its closure is drawn as a vertical narrow dashed line to avoid crossing the other contours.  The quantum-system contours remain open: we label the bra branch in replica $k$ by $i_k$ and the ket branch by $j_k$.  Thus, the diagrams do not represent the R\'enyi entropy itself; rather, they encode the time evolution of $\hat{R}_M$, the generalised density matrix of the $M$ replicated system.

\begin{figure}[h]
    \centering
    \includegraphics[width=0.5\linewidth]{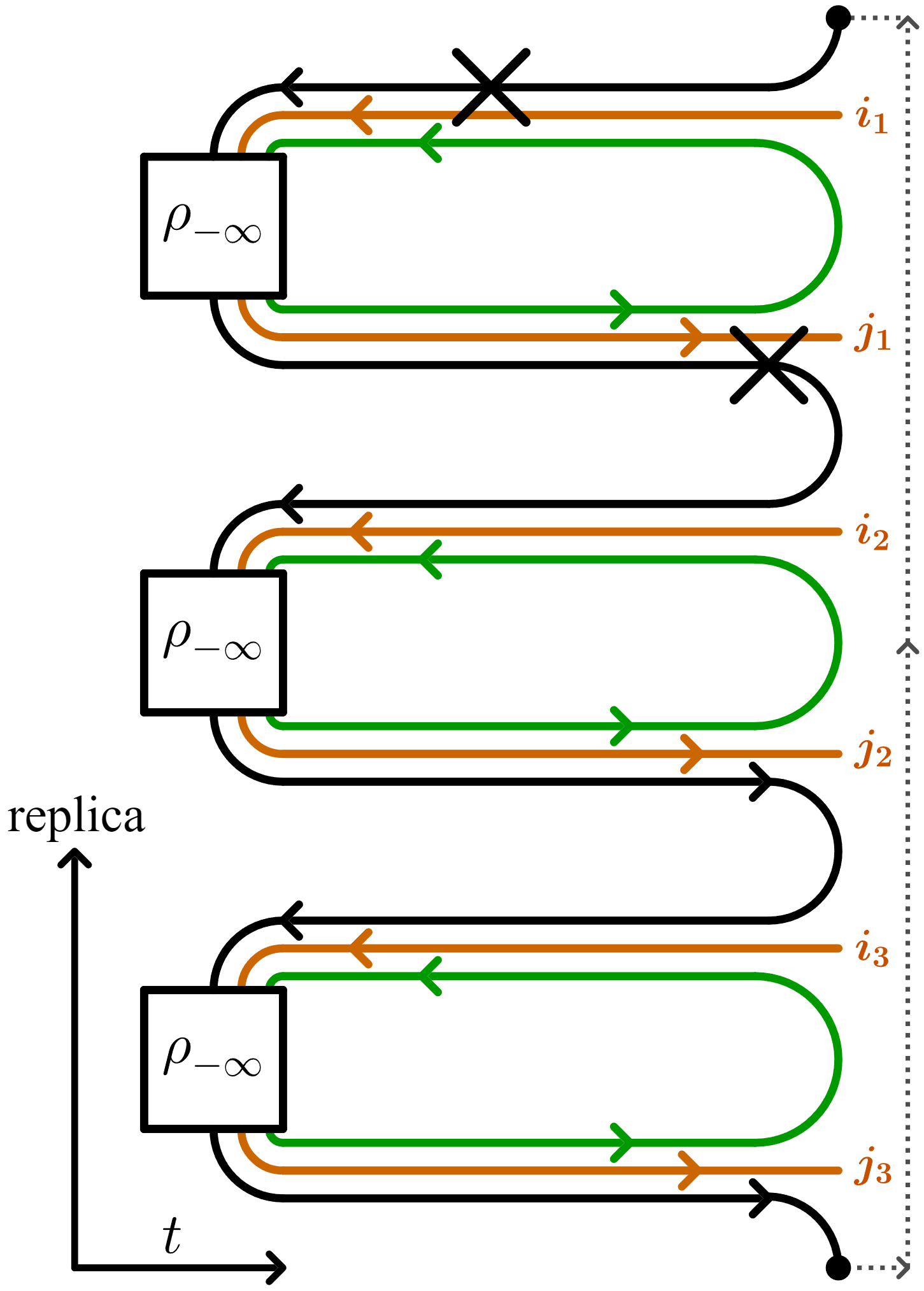}
    \caption{Diagrammatic representation of the \emph{intermediate-coupling} case.  As in the established KAN formalism, the replicas are linked through the probe-reservoir contours (black), which are closed only after encircling all replicas (periodic boundary condition), while the main reservoir (green) is traced out \emph{within} each replica.
    The quantum-system contour (orange) is left open inside every replica; its open ends carry explicit indices, allowing the entire construction to be collected into the generalized density matrix~\(R\).
    Crosses denote interaction vertices between the probe reservoir and the quantum system.   All contours terminate at the present time, but are straightened here for visual clarity.}
    \label{F: Open approach}
\end{figure}

\section{A Hybridized Qubit As Energy Valve}
\label{S: Qubit}

We will now apply the strong coupling formalism described above to a simple quantum system, which consists of an externally driven qubit. First let us describe the quantum system part. Let us consider the qubit frequency is $\hbar \omega= E_1-E_0$ and externally driven by amplitude $\hbar \Omega$ and the frequency $\omega_{\mathrm{dr}}$, making  the Hamiltonian
$\hat{H}_{dr}(t) = \hbar \Omega \cos(\omega_{dr}t) \left(|0\rangle \langle 1| +H.c. \right)$. The driving amplitude $\Omega$ is not that strong to inject new photon into the system, i.e. $\Omega \ll \omega$.

In order to make the driving part independent explicitly from time, we transform the Hamiltonian into a frame rotating with the driving frequency.  Performing the rotating wave approximation (RWA)  so that the Hamiltonian of the qubit subsystem becomes 
\begin{equation}
\label{E: RF Hamiltonian}
    \hat{H}_Q = -\frac{\delta}{2} \hat{Z} + \frac{\Omega}{2} \hat{X}
\end{equation}
with detuning $\delta = \omega- \omega_{dr}$.

Two reservoirs $e$ (environment) and $b$ (probe) are coupled to the quantum system. Linear response theory is considered to let the reservoirs interact only with the qubit. Each reservoir $R=\{e,b\}$ carries the free Hamiltonian  $\hat{H}_R = \sum_{k} \omega_k \hat{a}_{k}^\dagger \hat{a}_{k}$, with $\hat{a}_{k}$ denoting the ladder annihilation operator in the reservoir $R$ and the frequency mode $k$. In other words we consider each reservoir can be represented by a complete set of oscillators at a continuous frequency.  The spectrum of reservoir frequencies $\omega_k$ is assumed to be sufficiently dense to be considered continuous. 

The quantum system is coupled to the reservoirs R by individual oscillators in it by the following interaction Hamiltonian $H_{QR}=  
\sum_{ij}|i\rangle\langle j|\hat{X}^{(R)}_{ij}$, with $\hat{X}_{ij}^{(R)}$
being the operators in the space of reservoir and defined as  $\hat{X}^{(R)}_{ij} = \sum_k c^{(R)}_{ij,k} \hat{a}_k e^{-i\omega_k t} + H.c.$ The operator $\hat{a}_k$ being annihilation operator of mode frequency $\omega_k$. The coupling strength between the $0\leftrightarrow 1$ transition in the qubit and $i \leftrightarrow j$  transition of the mode $k$ determines the complex number $c^{(R)}_{ij,k}$. In fact this coefficient is important quantity as it determines the strength of the coupling between qubit and  reservoir as shown by connectors of their degrees of freedom in Fig. (\ref{fig.diag}). Throughout we consider driving frequency matches with the qubit frequency, i.e. $\delta=0$. Rotating the interaction of quantum system with each classicla system of reservoir transforms to the following 
\begin{equation}
\label{E: Coupling term 2}
\begin{split}
    \hat{H}_{QR} &= e^{i\omega_{dr} t}  \hat{\sigma}^\dagger \hat{X}^{(R)}_{10}(t) + e^{-i\omega_{dr} t} \hat{\sigma}  \hat{X}^{(R)}_{01}(t)\\
\end{split}
\end{equation}

We assume a linear response of each environment to the state of the quantum system. In this case, each environment is completely characterized by the set of frequency-dependent generalized dynamical susceptibilities
$\chi^{(R)}_{ij.pq}(\nu)$. 

\paragraph*{Dynamical susceptibility.}---The dynamical (retarded) susceptibility $\chi_{AB}(\omega)$ quantifies how fast and strong an observable $A$ under external driving field $f(t)$ is perturbed into $\delta\langle A(t)\rangle$ as the result of the influence of another observable $B$ at earlier time. In the linear-response theory, this can be written as 
$\delta\langle A(t)\rangle=\!\int_{-\infty}^{\infty}\!dt'\,
            \chi_{AB}(t-t')\,f(t')$
            with the dynamical susceptibility being $\chi_{AB}(t)=\frac{i}{\hbar}\,\Theta(t)\,
             \bigl\langle[A(t),B(0)]\bigr\rangle_{\!\mathrm{eq}}$ with the index $eq$ indicating the equilibrium fluctuations.

The dynamical susceptibility is related to the correlators of {$\hat{X}^{(R)}$ (for brevity here we drop the index $(R)$) defined as $S_{ij,pq}(\tau) \equiv \textup{Tr}_R \{\hat{X}_{ij}(0)\hat{X}_{pq}(\tau)\rho_R\}_{\!\mathrm{eq}}$}. Note that the correlator depends on the time delay $\tau$ between the two operators $X$' acting in the reservoir, with the time delay being minimally zero. We need to evaluate the half-sided Fourier transform of the correlator between the two intervals: $ \int_0^\infty d\tau \exp({i\omega \tau}) S_{ij,pq} (\pm \tau)  \equiv \frac{1}{2} S_{ij,pq} (\pm \omega) \pm i\Pi_{ij,pq} (\pm \omega)$, which plays a crucial role in the evaluation, especially due to its application in the Kubo-Martin-Schwinger \cite{KMS} and its generalization \cite{AnsariNazarovEngines} that relates the correlator to the dynamical susceptibilities $\tilde{\chi}_{ij,pq}(\nu)$ of the reservoir:
\begin{equation}
    S^{(R)}_{ij,pq}(\omega) =  n_B\left(\frac{\omega}{T_R}\right) \ \tilde{\chi}^{(R)}_{ij,pq}(\omega)
\end{equation}
with  $n_{B}(x)=1/(e^{x}-1)$, $T_{R}$ the reservoir temperature.

\section{Two-Replica Hybridization}
\label{S: Two Replicas}

To illustrate a practical construction, we restrict ourselves to the simplest non-trivial case, \(M = 2\): in two replicas a single qubit couples to two independent reservoirs.   
The Liouvillian super-operator \(\mathcal{L}\) decomposes naturally
\begin{equation}
  \mathcal{L} \;=\; \mathcal{L}_{U} \;+\; \mathcal{L}_{e} \;+\; \mathcal{L}_{b},
  \label{eq. total Liu}
\end{equation}
into three transparent contributions: (i) the unitary intrinsic evolution, \(\mathcal{L}_{U}\); (ii) the dissipative coupling to the main reservoir \(e\), \(\mathcal{L}_{e}\); and (iii) the interaction with the probe reservoir \(b\), \(\mathcal{L}_b\).

For \(M = 2\), the generalized density operator introduced in~ Eq. (\ref{E:Rhat_definition}) carries four indices--two for the kets and two for the bras of the replicas--so that
\[
  \hat{R}_{2} \;\equiv\; R_{i_{1} i_{2},\, j_{1} j_{2}}\,
  |i_{1} i_{2}\rangle \langle j_{1} j_{2}|.
\]
In Fig.~(\ref{F: Open approach}) this double-replica state is depicted as a red open contour.  In practise, we consider 2 two-level qubits, each one is described by $2\times 2$ matrix, which is doubly replicated and therefore the generalized density matrix has $4\times4$ arrays. We vectorize the density matrix into a $1\times 16$ vector and therefore in Eq. (\ref{E: Liouville equation}) the Liouvillian superoperator is a $16\times 16$ matrix containing all interaction in each replica and in between them. 

Each qubit replica (orange) couples to its local environment through the interaction Hamiltonian \(H_{RQ}\). At second order, two crosses denote the two interactions \(H_{RQ}(t_{1})\) and \(H_{RQ}(t_{2})\) between the qubit \(Q\) and the  reservoir R.  
Mixed terms involving a single \(RQ\) only in a reservoir vanish.  Comprehensive derivations of these diagrammatic rules---and their efficient resummations---can be found in  Refs.~\cite{AnsariNazarov16,AnsariNazarovEngines,AnsariSteenselNazarov}.

We summarize each term below.

\paragraph*{(i) Unitary part.}---The coherent evolution, dictated by the system Hamiltonian $\hat{H}_{Q}$, acts on the replicated density operator $\hat{R}_{2}$ through the commutator
\begin{equation}
\mathcal{L}_{U} \equiv -i\,[\hat{H}_{Q},\hat{R}_{2}] .
\label{eq:Unitary}
\end{equation}

\paragraph*{(ii) Coupling to the environment.}---Tracing out the modes of reservoir $e$ within each replica yields the canonical contribution \cite{doi:10.1142/S1230161217400017}:
\begin{equation}
\label{E: Main reservoir part}
\begin{split}
    \mathcal{L}_e =& \sum_k \left\{-i\Delta_e \comm{\hat{\sigma}_k^\dagger \hat{\sigma}_k}{\hat{R}^{(e)}_2}\right. \\
    & + \Gamma_{e\uparrow} \left( \hat{\sigma}_k^\dagger \hat{R}^{(e)}_2 \hat{\sigma}_k - \frac{1}{2} \left\{ \hat{\sigma}_k \hat{\sigma}_k^\dagger , \hat{R}^{(e)}_2 \right\} \right)\\
    & \left.+ \Gamma_{e\downarrow} \left( \hat{\sigma}_k \hat{R}^{(e)}_2 \hat{\sigma}_k^\dagger - \frac{1}{2} \left\{ \hat{\sigma}_k^\dagger \hat{\sigma}_k , \hat{R}^{(e)}_2 \right\} \right)\right\}
\end{split} 
\end{equation}
with $\Delta^{(2)}_{e}$ denotes the Lamb shift induced by the environment and defined as follows  
$\Delta_e = \Pi_{01,10}^{(0,1)}(\omega)-\Pi_{10,01}^{(0,1)}(-\omega)$. The transition rates $\Gamma_{e\uparrow}= \Gamma_{e\downarrow}\,\exp({-\omega/T_{e}})
= \chi_{_{e}}\,n_{B}({\omega}{T_{e}})$ and $\chi_{_{e}}$ its dynamical susceptibility of the environment.  The definition of $\Pi_{01,10}^{(0,1)}$ follows the generalized definition of correlators and the generalized KMS theorem on $M$ replica,  discussed in details and derived diagrammatically within the KAN formalism~\cite{AnsariNazarov16}.

\paragraph*{(iii)---Contribution of the probe reservoir.}---This part is a subtle yet indispensable component because its Keldysh contour is shared by both replicas.  This shared contour entangles the replicas at the level of perturbation theory and, consequently, demands a careful second–order treatment.

The expansion to quadratic order in the system–probe coupling reveals precisely two distinct diagrammatic topologies: (1) \emph{Same‑world diagrams}, where both interaction vertices reside in the contour of a single replica; (2) \emph{Cross‑world diagrams}, where an interaction takes place in each replica, thus explicitly linking their dynamical histories.

This classification streamlines the subsequent derivation and mirrors the physical intuition that the probe reservoir can act locally (within one replica) or non‑locally (between replicas) in Liouvillian space.

\begin{figure}[h]
\includegraphics[width=.18\textwidth]{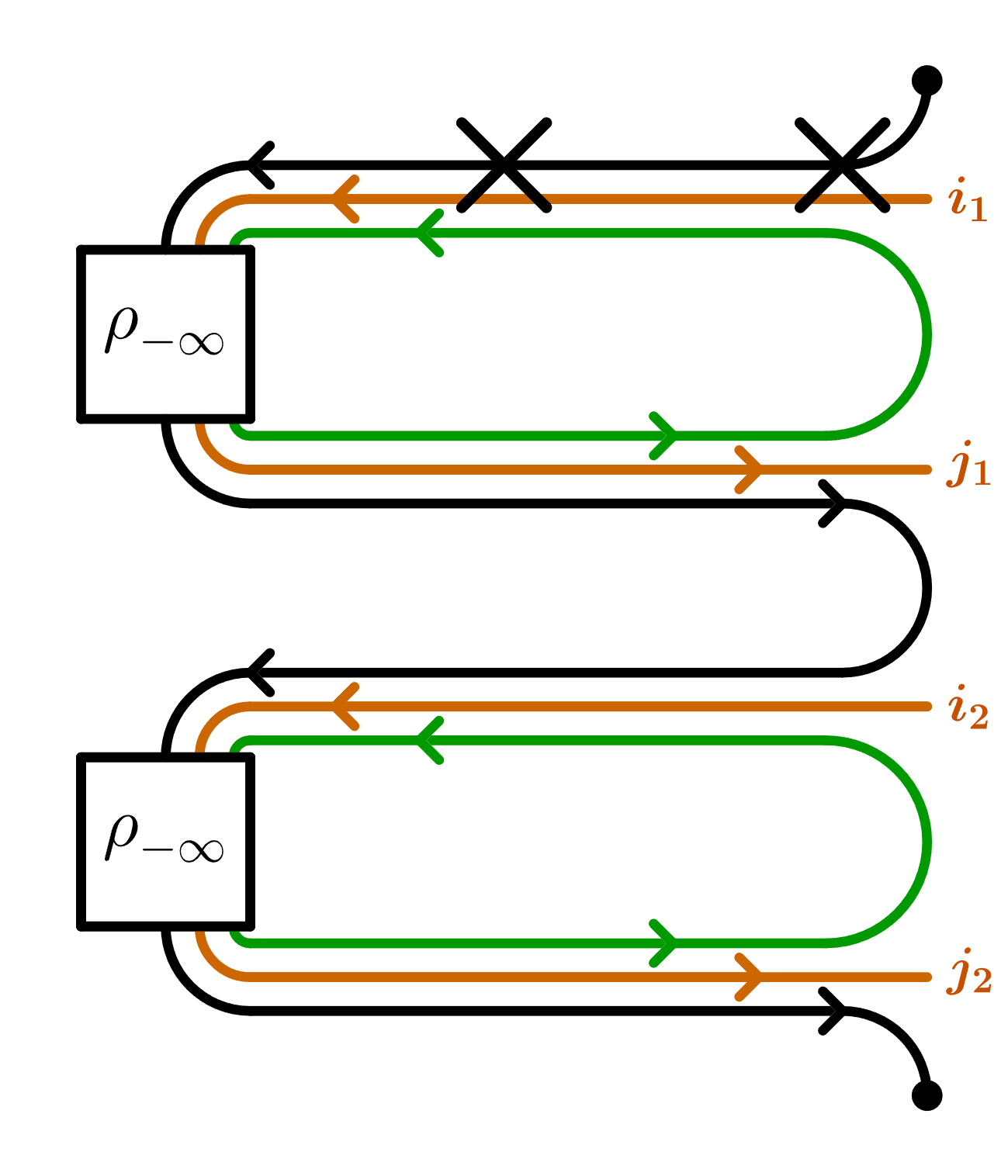}\hfill
    \includegraphics[width=.18\textwidth]{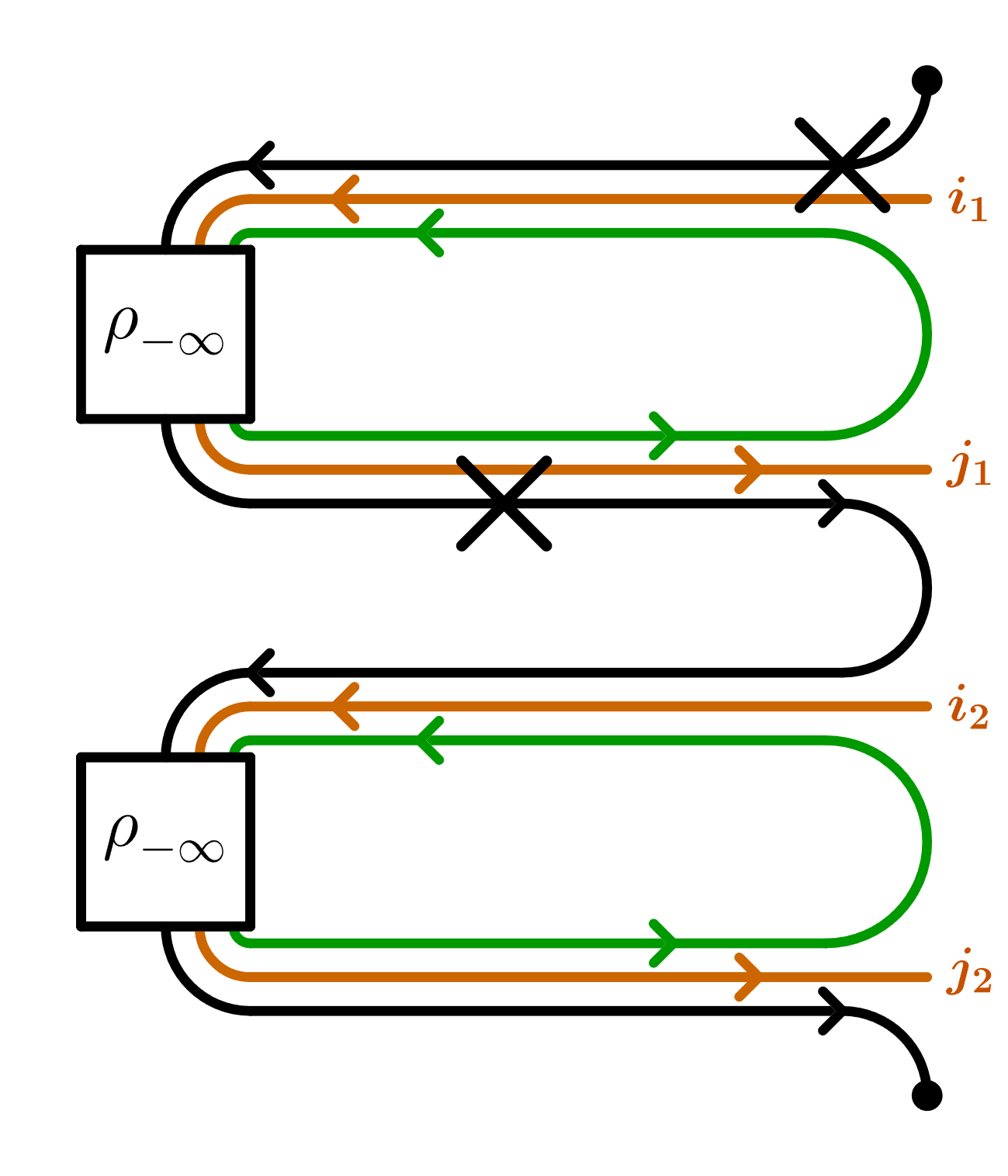}
    \\[\smallskipamount]
    \includegraphics[width=.18\textwidth]{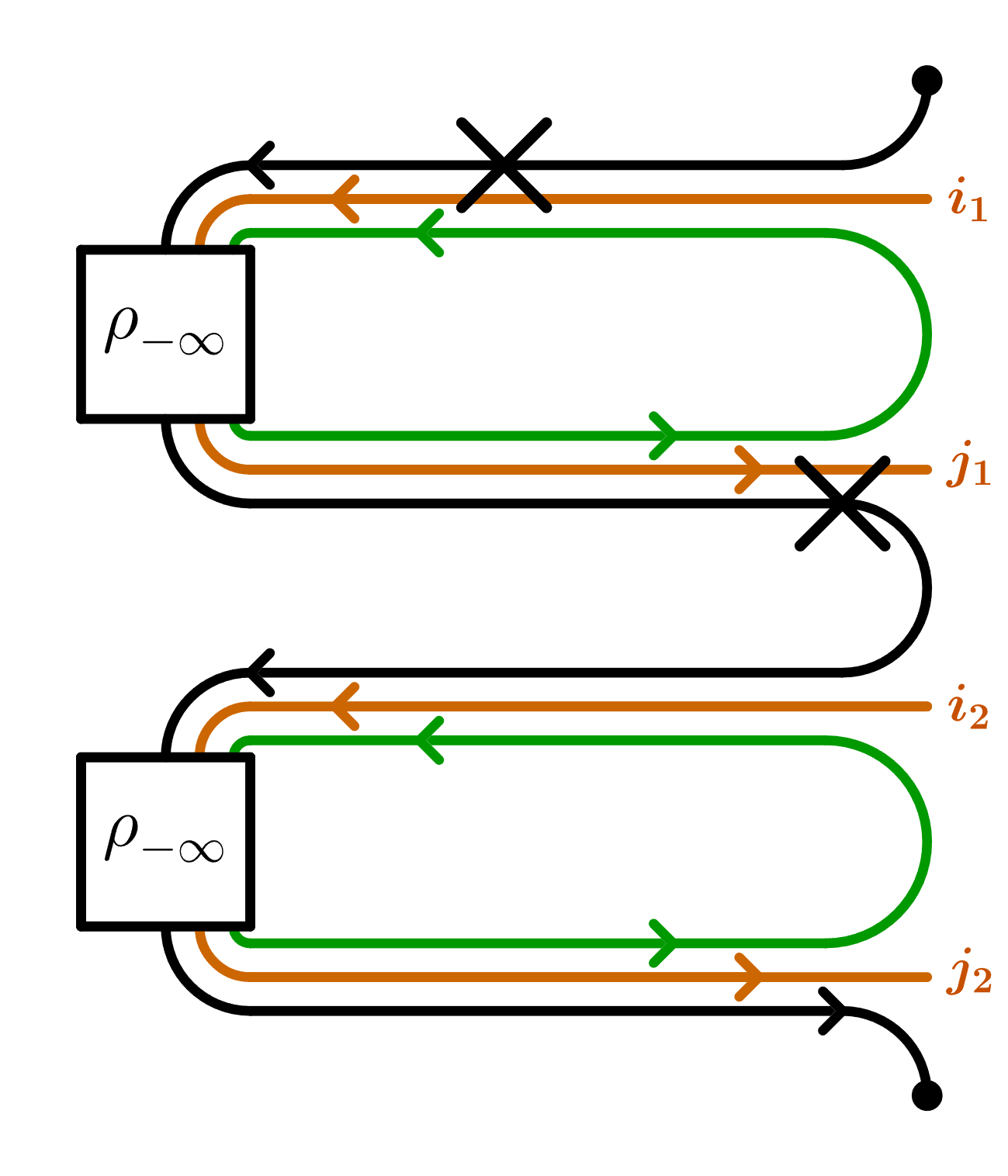}\hfill
    \includegraphics[width=.18\textwidth]{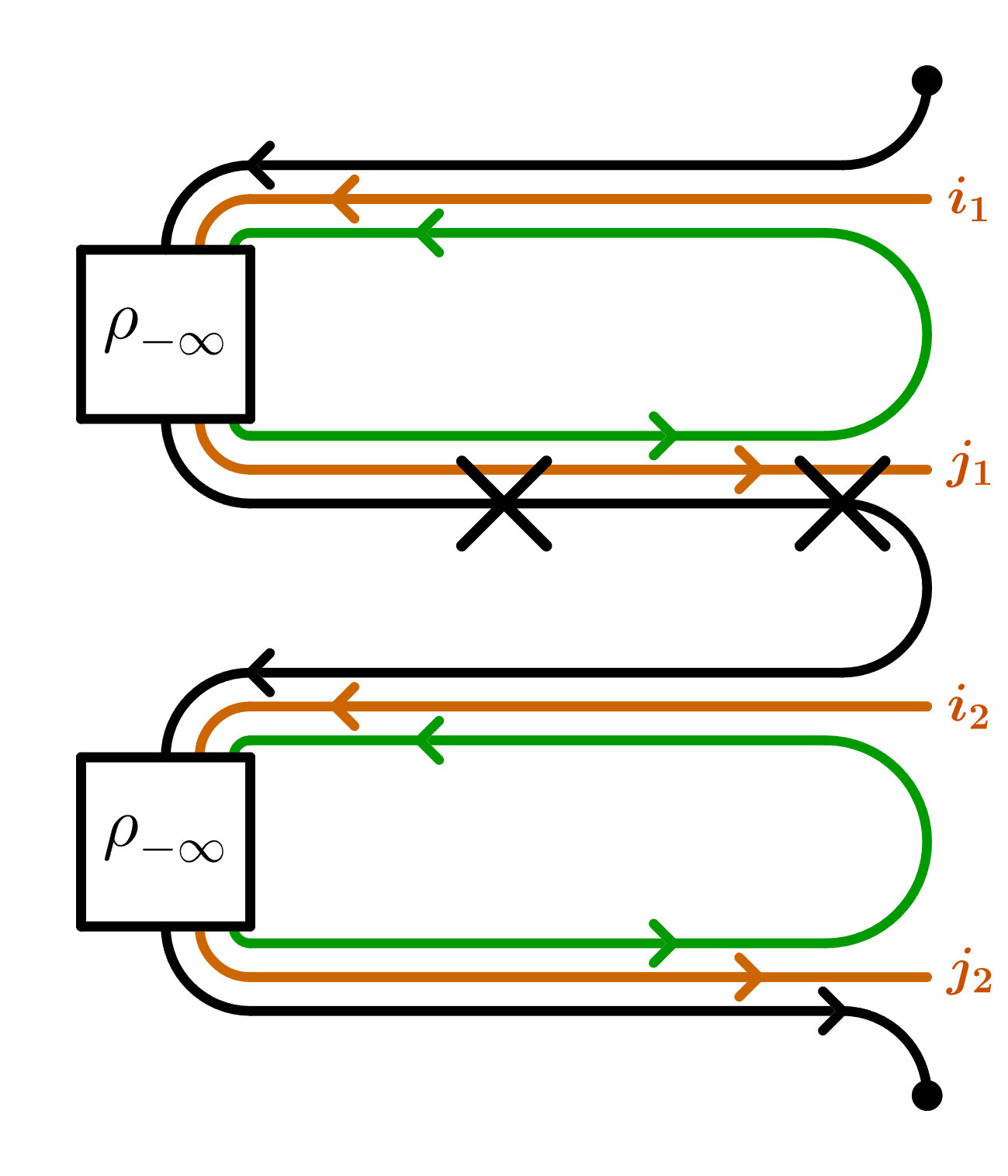}
    \caption{Four typical diagrams contributing to the same-world Liouvillian matrix for the case of two replicas. The later interaction is fixed at the present time $t$, while the other one is at some earlier time $t'$.}
    \label{F: Same-world diagrams}
\end{figure}

\begin{figure}[h]
    \centering
    \includegraphics[width=.18\textwidth]{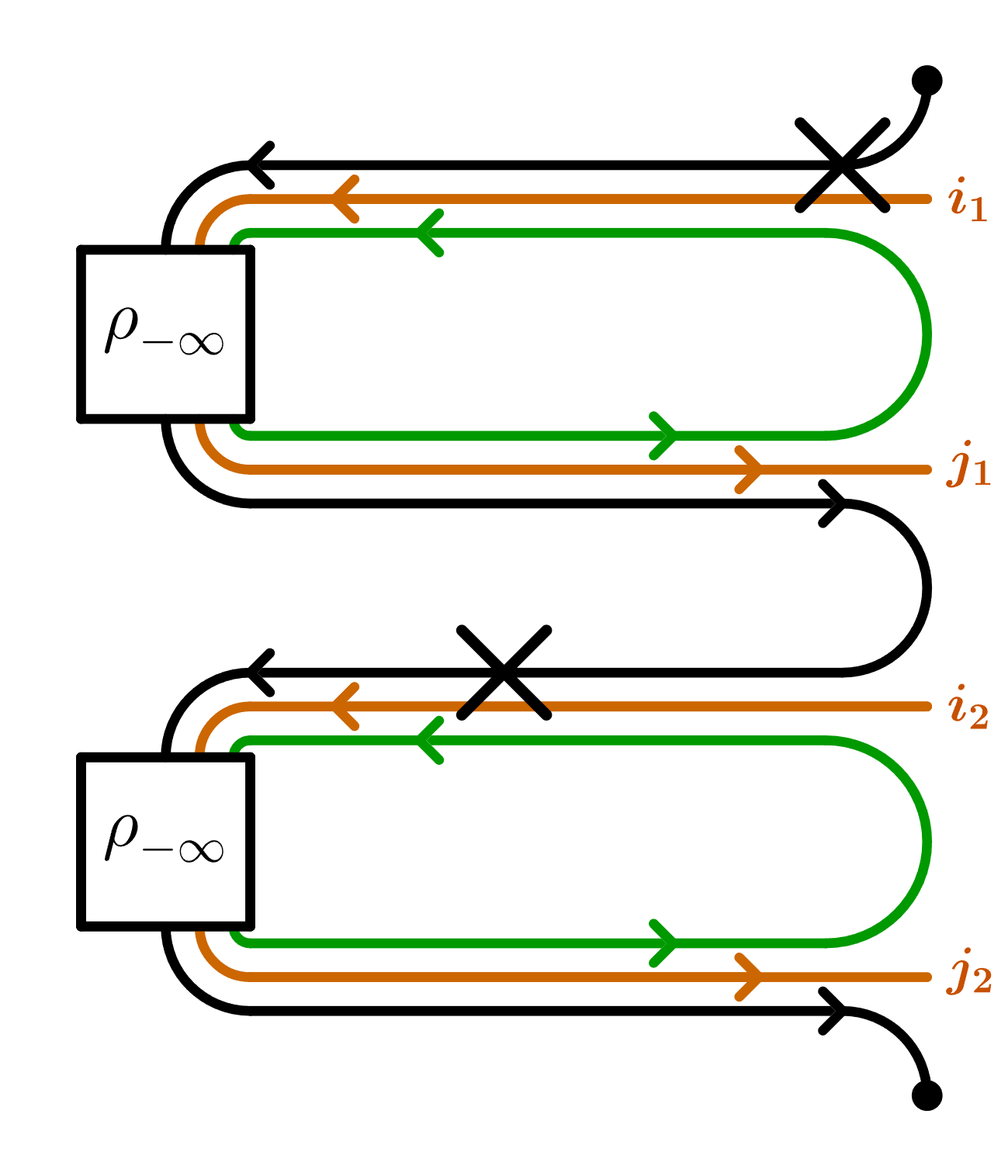}\hfill
    \includegraphics[width=.18\textwidth]{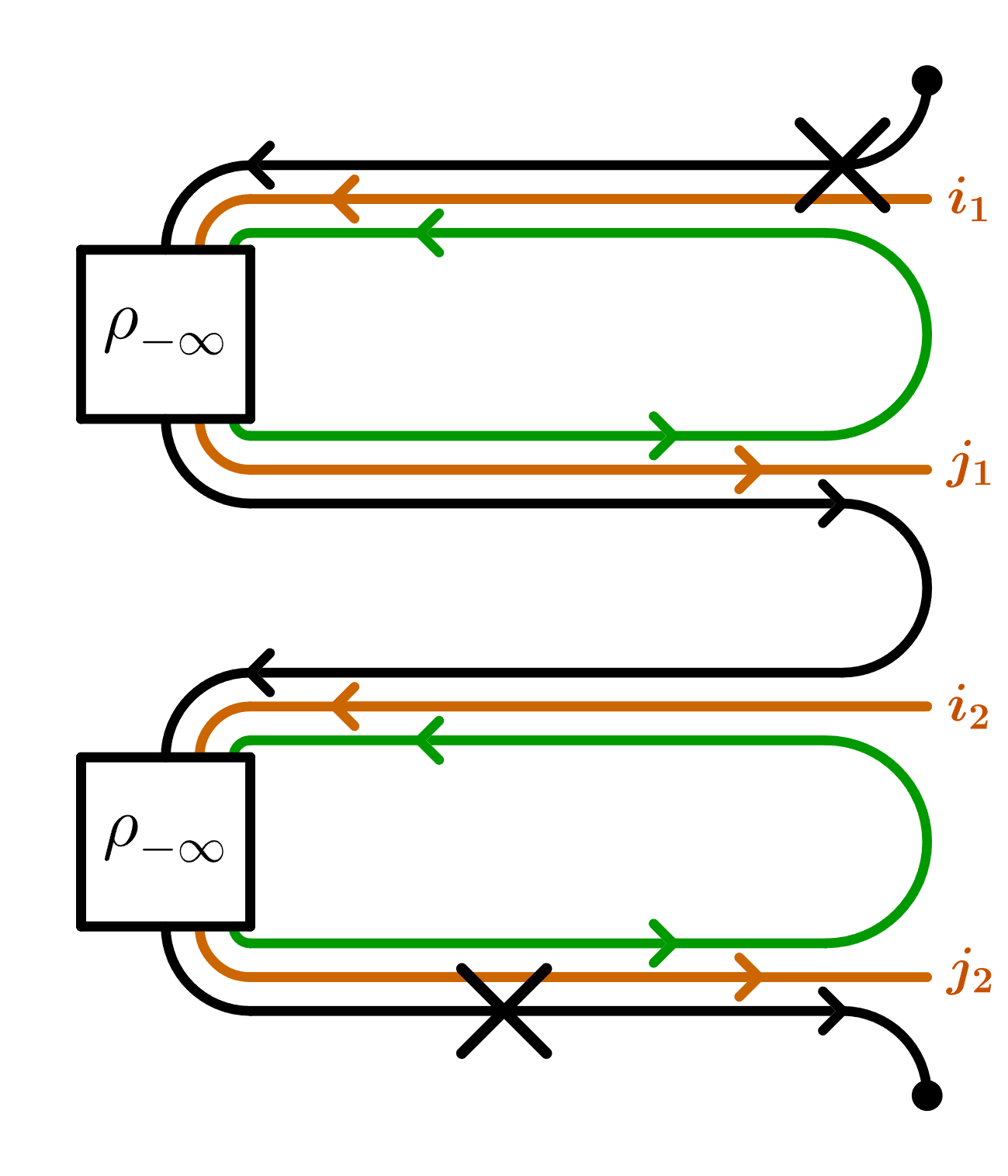}
    \\[\smallskipamount]
    \includegraphics[width=.18\textwidth]{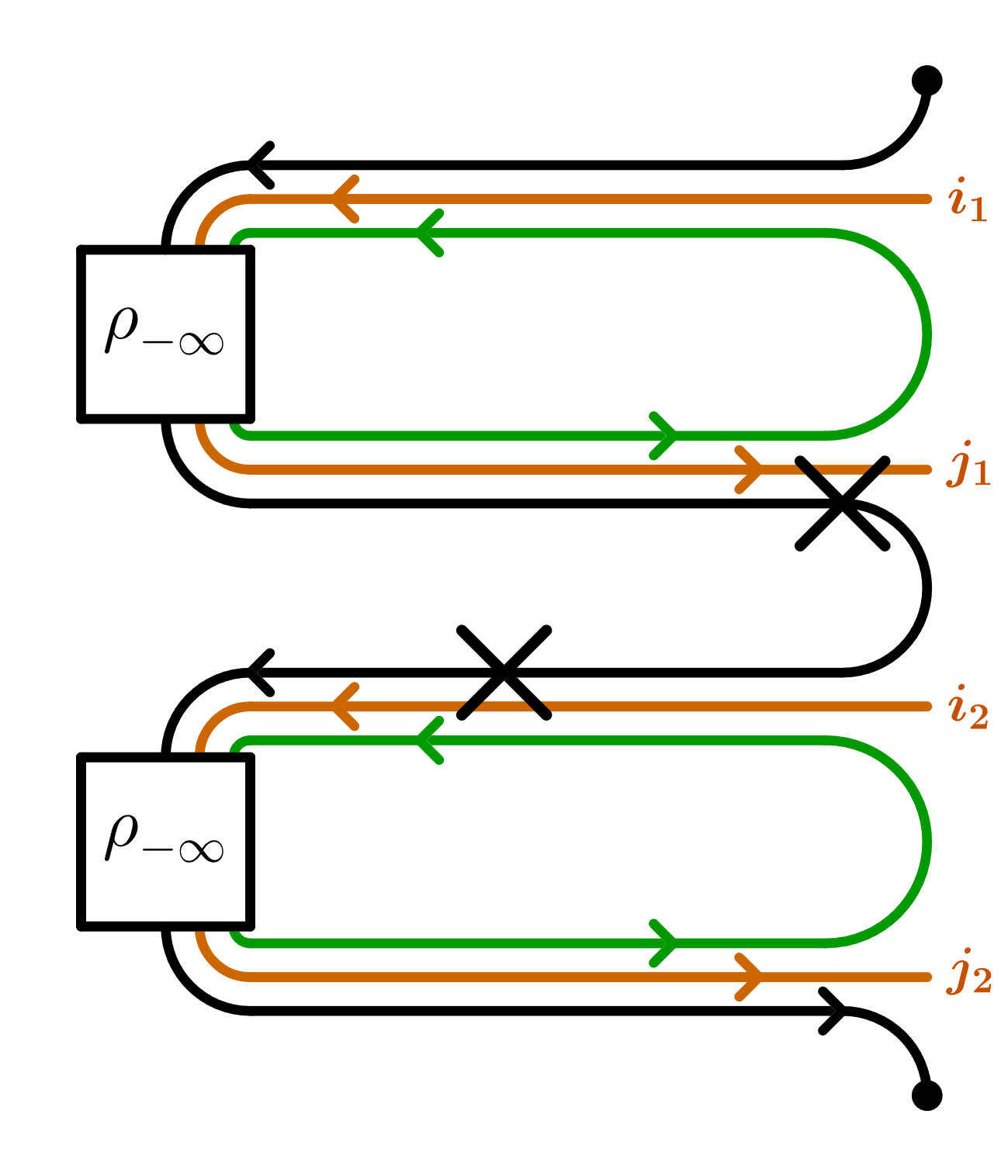}\hfill
    \includegraphics[width=.18\textwidth]{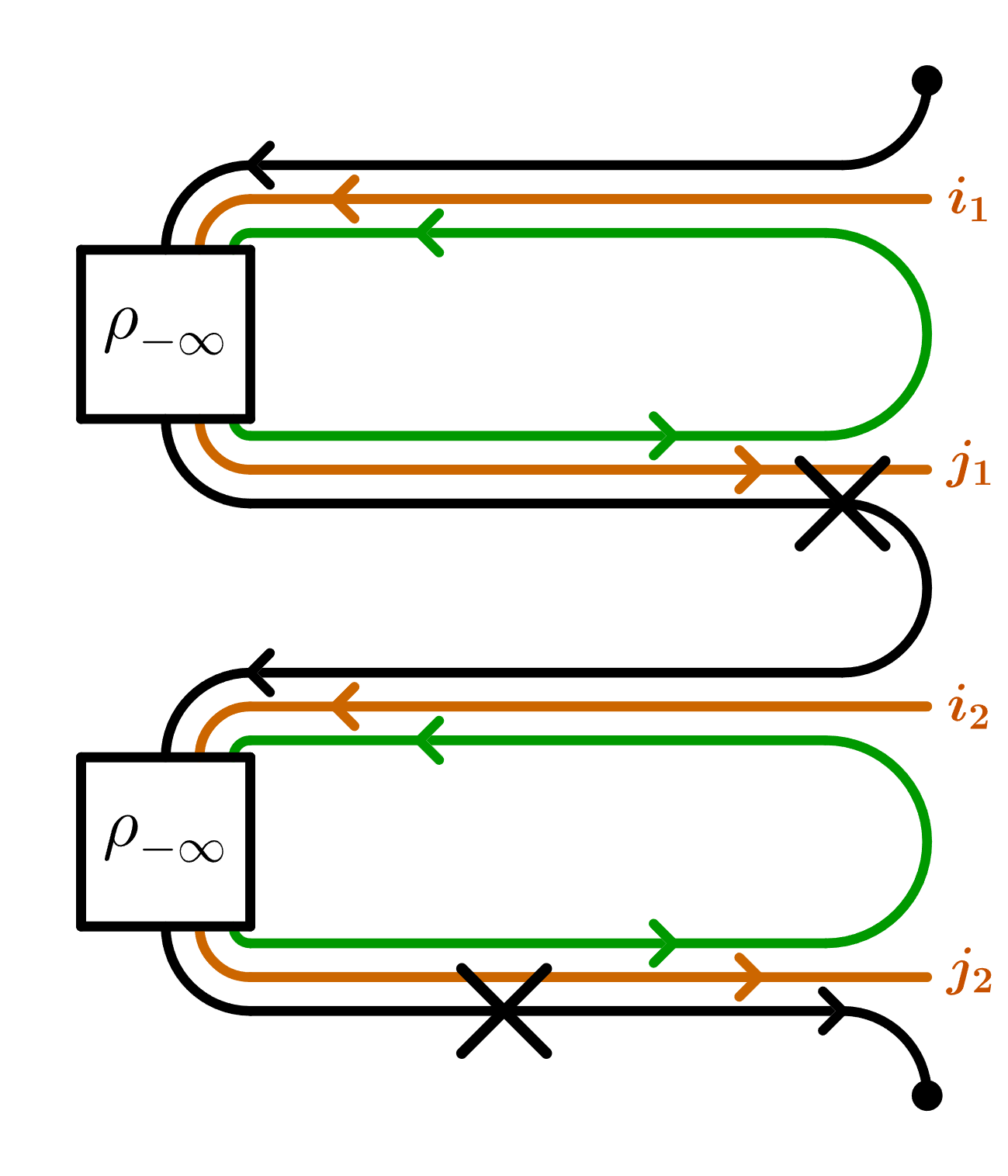} 
    \vspace{-0.3cm}
    \caption{Four out of eight diagrams contributing to the cross-world superoperator $\mathcal{C}_{12}$ for the case of two replicas, the other four diagrams are given by the time inversion $t \to t'$.}
    \label{F: Cross-world diagrams}
\end{figure}

\begin{figure*}[t]
\centering
    \includegraphics[width=.38\textwidth]{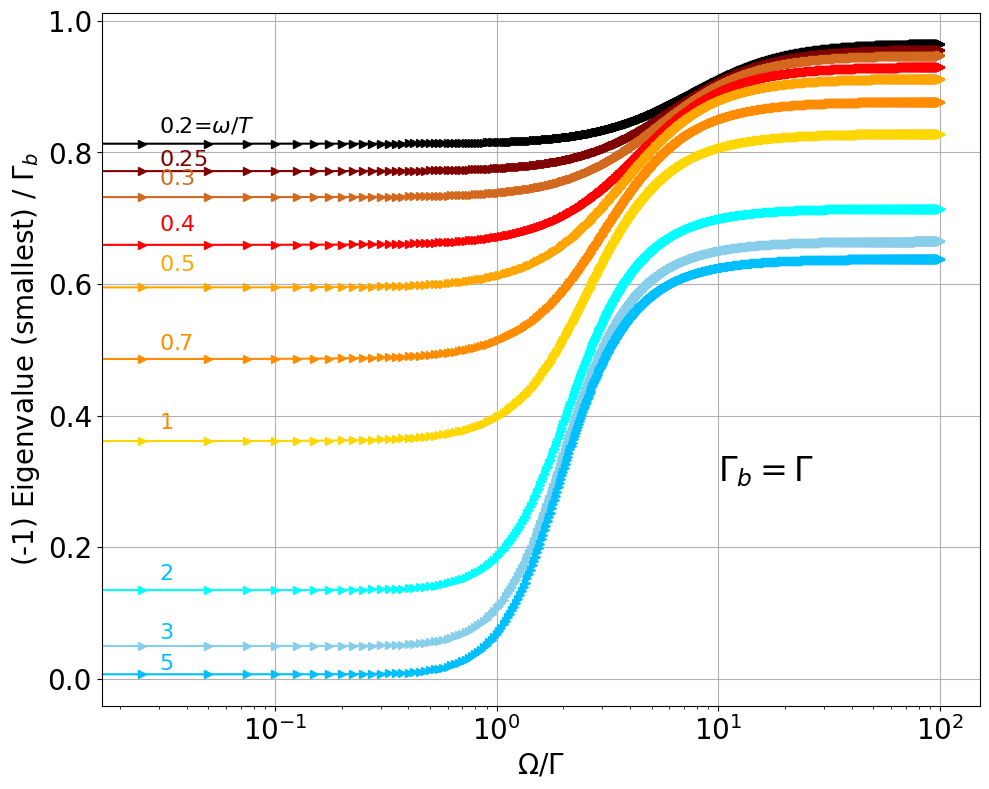}\put(-170,143){\textbf{(a)}}
\includegraphics[width=.38\textwidth]{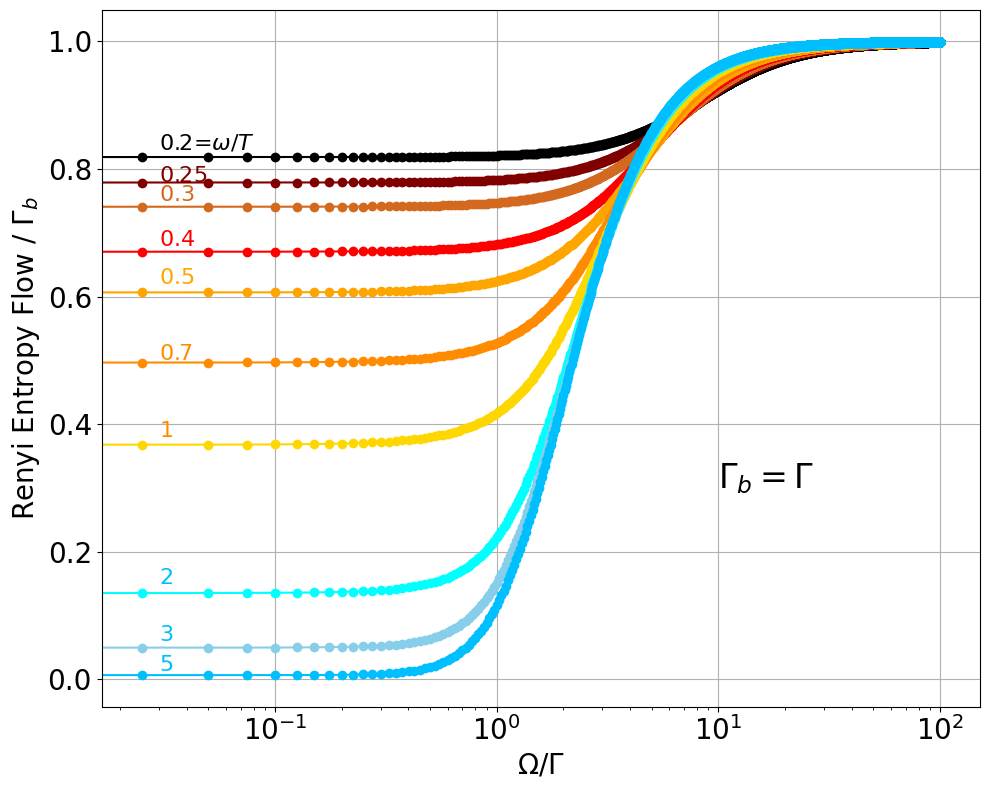} \put(-170,143){\textbf{(b)}}\\
\includegraphics[width=.38\textwidth]{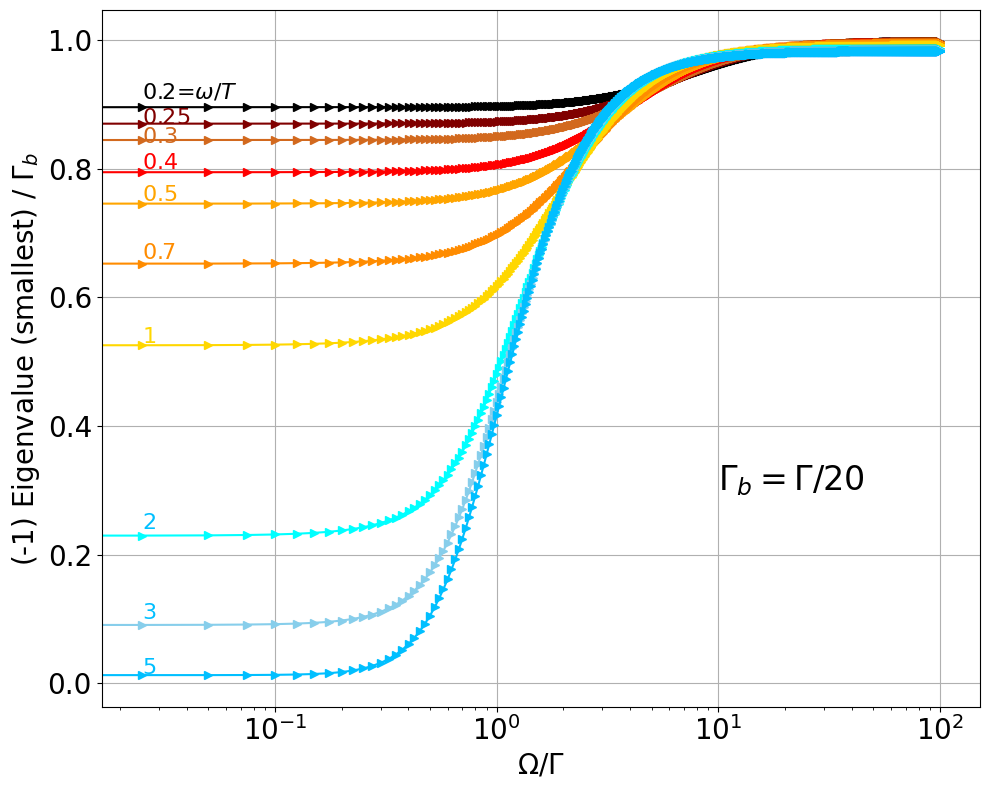} \put(-170,143){\textbf{(c)}}
\includegraphics[width=.38\textwidth]{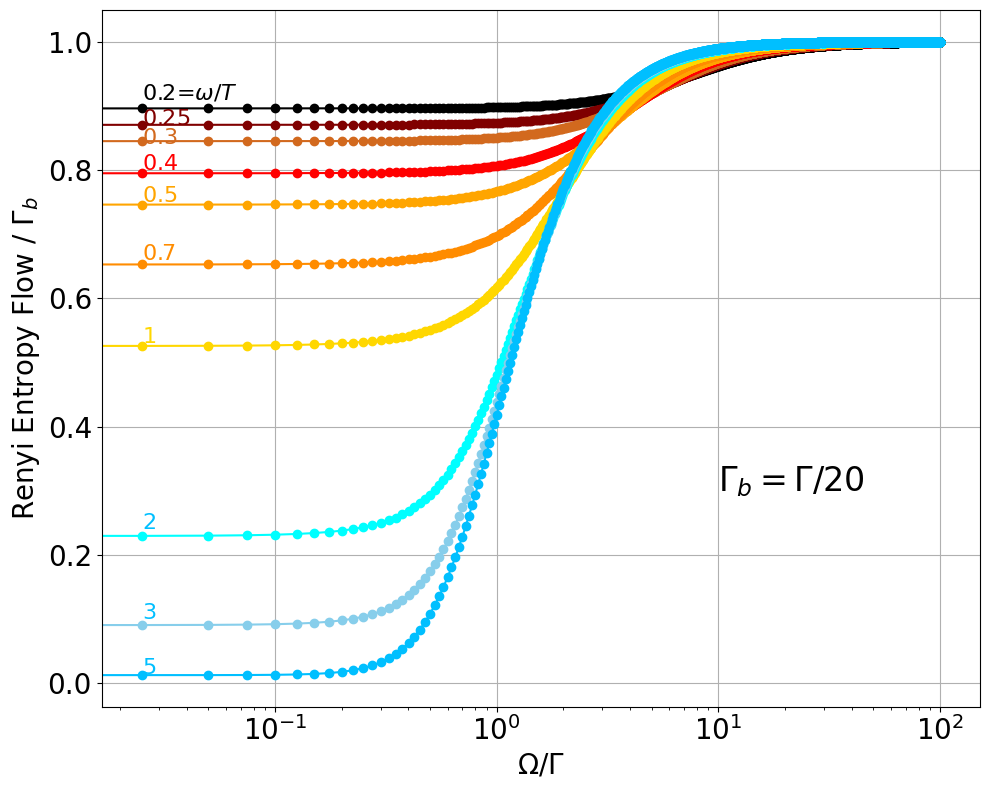} \put(-170,143){\textbf{(d)}}
    \caption{Entropy flow versus drive amplitude.  
Numerically computed probe R\'enyi entropy current $\dot S_2$ plotted against the reduced drive $\Omega/\Gamma_b$ for different choices of $\omega/T$.   The results in (a) and (c) follow the hybridization expression of Eq.~(\ref{E: Renyi conjecture}); while  (b) and (d) show the standard weak-coupling result of Eq.~(\ref{eq.Renyiflow}).  
Panels (a) and (b) correspond to strong probe coupling ($\Gamma_b=\Gamma_e$), while (c) and (d) illustrate the weak-coupling limit ($\Gamma_b/\Gamma_e\ll1$).}
    \label{F: Result comparison}
\end{figure*}

\paragraph*{Same‑world sector.}---Summing the diagrams of Fig.~(\ref{F: Same-world diagrams}) and the integration over the internal time coordinate yield the following Liuvillian matrix:
\begin{equation}
\label{E: Two replica same-world superoperator}
\begin{split}
    \mathcal{L}_{b,\mathrm{same}} =& \sum_{r=1,2} -i\Delta_{b}^{(r,2)}\left[\hat{\sigma}_{r}^{\dagger}\hat{\sigma}_{r},\hat{R}^{(b)}_2 \right]\\
    & + \Gamma_{ b \downarrow }^{(r,2)} \left( {e^{- \omega/T_b }} \, \hat{\sigma}_{r}^{\dagger} \hat{R}^{(b)}_2 \hat{\sigma}_{r}-\frac{1}{2}\left\{ \hat{\sigma}_{r}^{\dagger}\hat{\sigma}_{r}, \hat{R}^{(b)}_2 \right\} \right)\\
    & +\Gamma_{b\uparrow}^{(r,M)}\left( {e^{ \omega/T_b} } \hat{\sigma}_{r} \hat{R}^{(b)}_2 \hat{\sigma}_{r}^{\dagger}-\frac{1}{2}\left\{ \hat{\sigma}_{r} \hat{\sigma}_{r}^{\dagger}, \hat{R}^{(b)}_2 \right\} \right)
\end{split}
\end{equation}
with $r$ denoting the replica lanbel, $r=1,2,\cdots,M$. The multi replica Lamb shift and generalized transition rates are expressed in terms of the reservoir correlation functions,
\begin{equation}
\label{E: Multi-replica delta and gamma}
\begin{split}
    \Delta_{b}^{(r,M)} &= \Pi_{01,10}^{(0,M)}(\omega_{k})-\Pi_{10,01}^{(0,M)}(-\omega_{k}),\\
    \Gamma_{b\uparrow}^{(r,M)} &= \Gamma_{b\downarrow}^{(r,M)} {e^{-M  \omega/T_{b}}} = \chi_{_{b}} {n}_B(M \omega/ T_{b}).
\end{split}
\end{equation}

Eq. (\ref{E: Two replica same-world superoperator}) retains the standard Lindblad footprint, but the exponential prefactors $\exp({\pm\omega/T_{b}})$ reverse the intuitive functions of excitation and relaxation once the operator acts on the semiclassical diagonal of the density matrix, a direct manifestation of the nonequilibrium influence of the probe reservoir.  The same‑world contribution for the second replica is obtained by interchanging $\hat{\sigma}_1 \leftrightarrow \hat{\sigma}_2$.

\paragraph*{Cross‑world sector.}---As illustrated in Fig. (\ref{fig.repdiag}), there is no physical or mathematical principle that forbids considering the consistent scenario in which a particle is created in one replica and annihilated in another. This has no relation to notions such as a multiverse or higher dimensions, since we make no hypothesis about the physical existence of multiple universes in which both mathematical and physical quantities are evaluated. Instead, all physical quantities are ultimately evaluated within a single replica, after analytically continuing the number of replicas to one. More precisely, we employ a consistent replica trick to handle the nonlinearity of density-matrix functions appearing in informational measures such as entropy. The generalization is purely a mathematical device, and only after reducing the number of replicas to one do we interpret the results in physical terms.

Such cross terms have been depicted in diagrams of Fig. (\ref{F: Cross-world diagrams}) in which each replica contains only one interaction in time, when is summed over for all possible such exchanges will produce
\begin{equation}
\label{E: Two replica cross superoperator}
\begin{split}
    \mathcal{L}_{b,\mathrm{cross}} = \Gamma_{{b}\downarrow}^{(2)} \bigg( & \hat{\sigma}_{2} \hat{R}^{(b)}_2 \hat{\sigma}_{1}^{\dagger} + {e^{-2 \omega/T_{b} }} \hat{\sigma}_{1}^{\dagger} \hat{R}^{(b)}_2 \hat{\sigma}_{2}\\
    & - {e^{- \omega/T_{b}}} \left\{ \hat{\sigma}_{1}^{\dagger} \hat{\sigma}_{2}, \hat{R}^{(b)}_2 \right\} \bigg)\\
    +\Gamma_{{b}\uparrow}^{(2)} \bigg( & \hat{\sigma}_{2}^{\dagger} \hat{R}^{(b)}_2 \hat{\sigma}_{1} + {e^{2 \omega/T{b}}}\hat{\sigma}_{1} \hat{R}^{(b)}_2 \hat{\sigma}_{2}^{\dagger}\\
    & - {e^{ \omega/T_{b}}} \left\{ \hat{\sigma}_{1} \hat{\sigma}_{2}^{\dagger}, \hat{R}^{(b)}_2 \right\} \bigg)
\end{split}
\end{equation}

  \begin{figure*}[t]
    \centering
\includegraphics[width=0.7\linewidth]{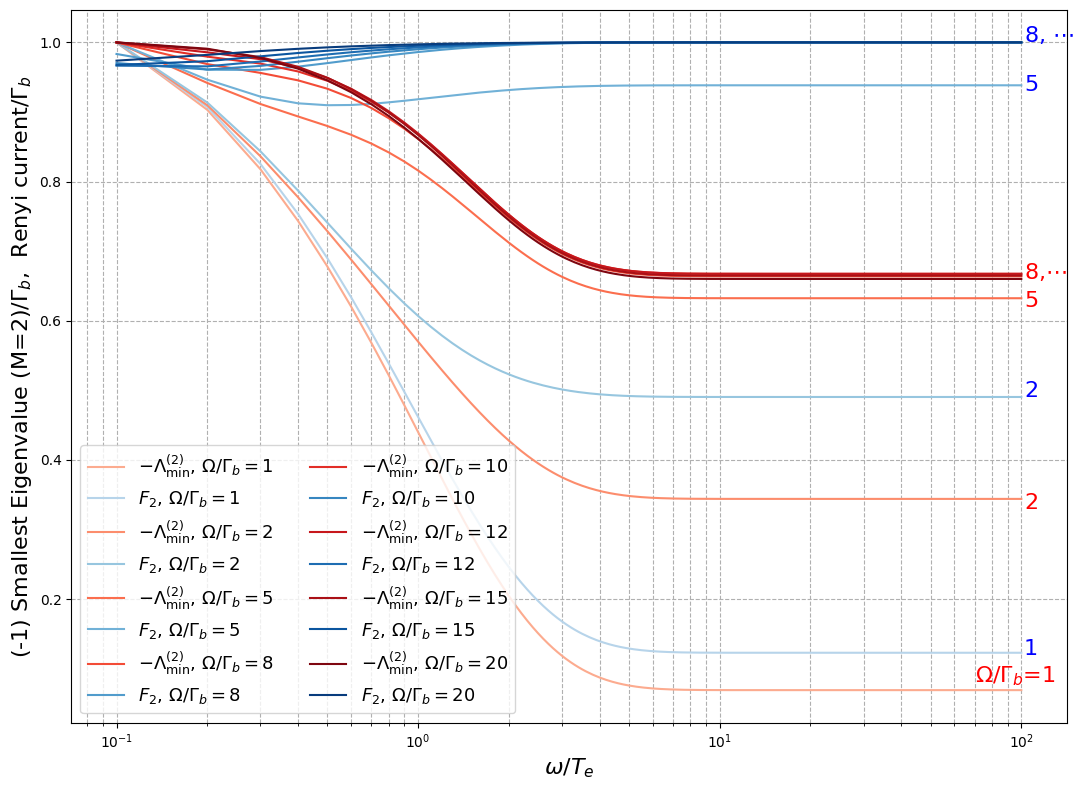}
    \caption{Comparison of entropy‐flow results in the strong‐coupling regime ($\Gamma_b = \Gamma$) between the hybridization formalism (red) and the standard weak‐coupling formalism (blue) as functions of $\Omega/T_e$ for various $\Omega/\Gamma_b$. The two approaches agree for weak driving but diverge markedly at higher drive amplitudes, revealing the significance of hybridization effects.}
    \label{fig:compare}
\end{figure*}

No additional Lamb shift arises because the principal-value parts of each time-reversed pair cancel exactly.  Physically, this cross world encodes reservoir-mediated coherence transfer between replicas and constitutes the key innovation of the multi-replica framework. This is in fact the term that can result in the quantum effect of tunneling through replicas as described in Fig. (\ref{fig.repdiag}).

Collecting all terms yields the total Liouvillian matrix, a compact $16\times16$ matrix that governs the dynamics of the replicated density operator $\hat{R}_{2}$ and, via the replica limit, direct produces the R\'enyi-2 entropy production rate. The total Liouvillian superoperator for $M=2$ has been explicitly seen in Appendix (\ref{appendix. Liuv}).

  After determining the Liouvillian matrix, we construct an orthonormal supervector basis $\$|\Phi\rangle\!\rangle\}$ that diagonalizes the total Liouvillian superoperator. The entropy flow obtained from Eq.~(\ref{E: Renyi conjecture}) is the smallest Liouvillian eigenvalue.

 We numerically evaluate the steady‐state R\'enyi entropy current flowing into the probe reservoir over a broad range of drive amplitudes and probe couplings. Figure \ref{F: Result comparison} presents these results, comparing the two theoretical prescriptions employed throughout this work: the hybridization formula, Eq. \eqref{E: Renyi conjecture}, shown in panels (a) and (c), and the standard expression, Eq. \eqref{eq.Renyiflow}, shown in panels (b) and (d). In all panels, the horizontal axis represents the dimensionless drive $\Omega/\Gamma_b$, defined as the ratio of the coherent Rabi frequency $\Omega$ to the bare probe transition rate $\Gamma_b$. Each panel contains multiple curves corresponding to different values of $\omega/T_e$, i.e., the qubit frequency divided by the environment temperature. The calculations are performed assuming a very low probe-reservoir temperature, $\omega/T_b \ll 100$.

For weak coupling, the two formalisms yield consistent results, as seen by comparing panels (c) and (d) for $\Gamma_b = \Gamma/20$. In contrast, a clear discrepancy emerges in the strong‐coupling regime, $\Gamma_b = \Gamma$, illustrated in panels (a) and (b). Here, panel (a) shows results from the hybridization formulation, whereas panel (b) displays those obtained using the weak‐coupling expression.   

To examine the strong‐coupling regime $\Gamma_b = \Gamma$ in more detail, we directly compare the two formalisms by plotting them together in Fig. (\ref{fig:compare}). The figure shows the entropy‐flow results from the hybridization formalism, $\Lambda^{(2)}_{\min}$ (red curves), and from the standard formalism, $\mathcal{F}_2$ (blue curves), as functions of $\omega/T_e$ for various renormalized drive amplitudes $\Omega/\Gamma_b$. The red curves represent the hybridized results, while the blue curves correspond to the weak‐coupling approach; in each color set, darker shades indicate stronger drive amplitudes.

The two approaches agree in the absence of drive and for weak driving, but deviate significantly as the driving amplitude increases, highlighting the impact of hybridization effects in the strong‐coupling limit.

Overall, Fig.~\ref{F: Result comparison} highlights that significant quantitative differences in entropy production emerge once the probe is promoted from a passive weakly coupled detector to an active, strongly hybridized partner at lower temperatures where quantum physics dominates classical physics.

\section{M-Replica Case}
\label{S: M Replicas}

We now extend the replica construction from two copies to an arbitrary integer number $M$.  For each extra replica, the generalized density operator becomes a tensor acting on a Hilbert space of dimension $2^M$, so its Liouvillian super-operator acting on space of dinmension $2^{2M}$ and lives in a space of dimension $2^{4M}$.  As long as $M$ is an integer, all of the required linear-algebraic objects—operators, tensor products, traces—are perfectly well-defined.

Trouble
arises only when one tries to treat $M$ as a continuous parameter: extending these
matrices to non-integer dimensions is not meaningful, so $S_M$ is generally non-differentiable at non-integer $M$.  This lack of differentiability complicates the formal route to the von Neumann entropy (which would require $\partial_M S_M|_{M=1}$),  but it does not prevent us from constructing well-behaved physical models for every integer $M$.

\paragraph*{Same-world super-operators.}---For each replica the unitary evolution and the coupling to the principal reservoir are identical to the two-replica case; extra worlds merely contribute additional, mutually non-interacting ``same-world'' blocks in the Liouvillian.  By building the super-operator explicitly for successive integers $M=2,3,\dots$ and then analyzing how its eigenstructure changes, one can still extract reliable predictions for the von Neumann entropy production in the delicate limit $M\to1$.  In particular, focusing on a given world $i$, only the intra-world (same-world) term is modified by the presence of the other $M-1$ replicas, while all inter-world (different-world) dissipators remain unchanged.  This systematic construction allows the entropy production rate to be inferred
indirectly, even though a formal derivative of $S_M$ with respect to $M$ is
ill-defined outside the integers.

\begin{widetext}
\begin{equation}
\label{E: M replica same-world superoperator}
\begin{split}
    \mathcal{S}_i\left[\hat{R}_2^{(b)} \right] =& -i\Delta^{(b,M)}\left[\hat{\sigma}_{i}^{\dagger}\hat{\sigma}_{i},\hat{R}_2^{(b)} \right]
     + \Gamma_{\downarrow}^{(b,M)}\left(e^{-\beta_b} \omega\hat{\sigma}_{i}^{\dagger} \hat{R}_2^{(b)} \hat{\sigma}_{i}-\frac{1}{2}\left\{ \hat{\sigma}_{i}^{\dagger}\hat{\sigma}_{i}, \hat{R}_2^{(b)} \right\} \right)
+\Gamma_{\uparrow}^{(b,M)}\left(e^{\beta_b \omega} \hat{\sigma}_{i} \hat{R}_2^{(b)} \hat{\sigma}_{i}^{\dagger}-\frac{1}{2}\left\{ \hat{\sigma}_{i} \hat{\sigma}_{i}^{\dagger}, \hat{R}_2^{(b)} \right\} \right)
\end{split}
\end{equation}
\end{widetext}

\paragraph*{Cross-world super-operators.}---The {cross-world} super-operators---those that transfer excitations between two
distinct replicas---keep essentially the same algebraic form they had in the
two-replica model.  What {does} change is the thermal weight that multiplies
them.  Whenever a jump carries an excitation from world $i$ to world $i+N$,
the process must ``pass through'' the $N-1$ intermediate replicas, which remain
completely decoupled from the bath.  These silent worlds act as an imaginary-time
delay, so the generalized Kubo–Martin–Schwinger relation (\ref{E: Generalized KMS}) injects a Boltzmann factor.  Concretely, an annihilation (creation) operator that hops across $N$ gaps is multiplied by
$\exp\!\left[\pm\tfrac{1}{2}\beta N\hbar\omega\right]$
exactly as if it had been evolved by $\exp(i\omega (t+i\tau) )$ for an imaginary time
$\tau=\mp N\beta\hbar/2$.

Collecting these factors, the cross-world dissipator that links replica \(i\) to
replica \(i+N\) can be written schematically as $\mathcal{L}_{i\rightarrow i+N} = 
\exp\left({-\beta N\hbar\omega/2}\right) \mathcal{J}_{\!\downarrow}[\rho] + 
\exp\left({\beta N\hbar\omega/2}\right) \mathcal{J}_{\!\uparrow}[\rho]$, where $\mathcal{J}_{\!\downarrow}$ ($\mathcal{J}_{\!\uparrow}$) annihilates
(creates) a system excitation while feeding energy back to (extracting energy from) the bath.  Apart from the thermal prefactors that depend explicitly on the
number $N$ of non-interacting replicas between the two active worlds, every
structural element---commutators, projectors, and the overall Lindblad form---remains unchanged. Consequently, the eigenbasis in which these operators act and their combination with the intra-world terms carry over intact to the full \(M\)-replica construction.

\begin{figure*}[t]
    \centering
\includegraphics[width=0.315\textwidth]{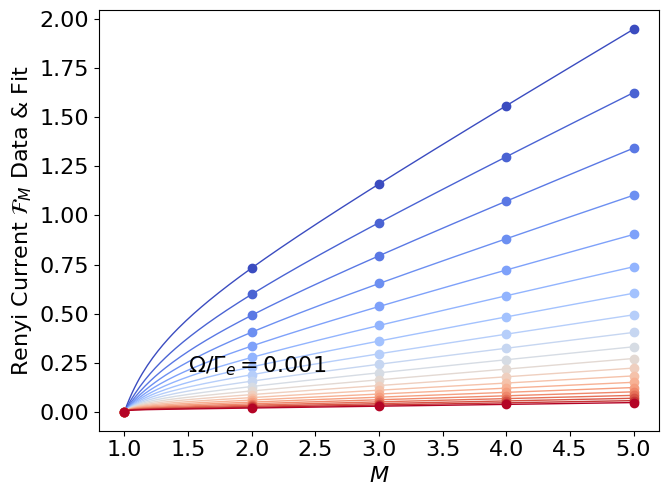} \put(-135,85){\textbf{(a)}}   \includegraphics[width=0.315\textwidth]{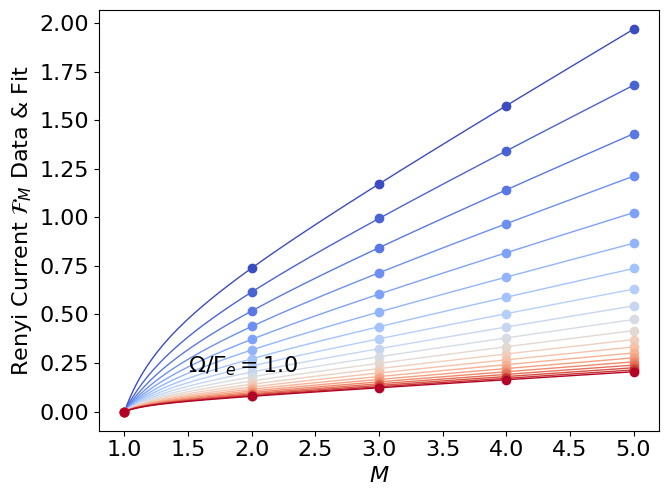}   \put(-135,85){\textbf{(b)}} 
\includegraphics[width=0.375\textwidth]{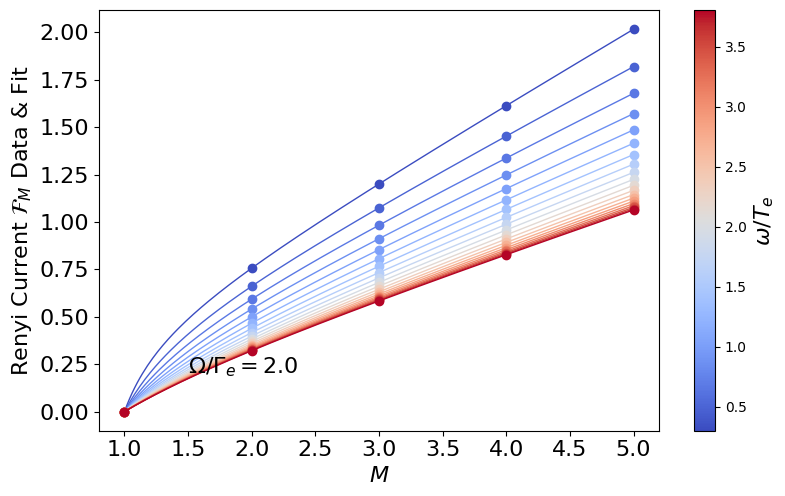}
   \put(-135,85){\textbf{(c)}}
    \caption{(a-c) Evaluation of R\'enyi entropy current of $M$ replicas versus $M$ (circle symbols) from strong coupling regime where the probe reservoir is coupled with the same coupling strength as the real encironment to the quantum system, i.e. $\Gamma_b=\Gamma_e$. Fitting function (solid lines) fills the flow for interpolating points associated with non-integer $M$. In (a) $\Omega/\Gamma_b=0.1$, in (b) 1 and in (c) 10.    (d-f) show the fitting parameters $a$, $b$, and $c$, respectively versus qubit frequency divided by the environment tempearture $\omega/T_e$,  for the three different values of weak (orange), intermediate (green), and strong driving amplitude (red). }
    \label{F: Lambda vs M}
\end{figure*}

\begin{widetext}
\begin{equation}
\label{E: M replica cross pair superoperator}
\begin{split}
    \mathcal{C}_{i,i+N} \left[\hat{R}_2^{(b)} \right]  =\, & \Gamma_{\downarrow}^{(b,M)} \Big( e^{-(N-1)\beta_b \omega} \hat{\sigma}_{i+N} \hat{R}_2^{(b)} \hat{\sigma}_{i}^{\dagger} 
     + e^{-(N+1)\beta_b \omega} \hat{\sigma}_{i}^{\dagger} \hat{R}_2^{(b)} \hat{\sigma}_{i+N} -e^{-N\beta_b \omega}\left\{ \hat{\sigma}_{i}^{\dagger} \hat{\sigma}_{i+N}, \hat{R}_2^{(b)} \right\} \Big)  \\ 
    + & \Gamma_{\uparrow}^{(b,M)} \Big( e^{(N-1)\beta_b \omega}\hat{\sigma}_{i+N}^{\dagger} \hat{R}_2^{(b)} \hat{\sigma}_{i} +e^{(N+1)\beta_b \omega} \hat{\sigma}_{i} \hat{R}_2^{(b)} \hat{\sigma}_{i+N}^{\dagger}   
    -  e^{N\beta_b \omega} \left\{ \hat{\sigma}_{i} \hat{\sigma}_{i+N}^{\dagger}, \hat{R}_2^{(b)} \right\} \Big) 
\end{split}
\end{equation}
\end{widetext}

\paragraph*{Gathering all contributions.}---Having identified the {intra-world} dissipators 
and the {cross-world} terms $\mathcal{L}_{i\!\rightarrow\!j}$ (transferring excitations from replica $i$ to
replica $j$), the complete generator of the open-system dynamics is obtained by a
straightforward super-operator sum.  Because each replica couples locally to the bath
and the replicas interact only through the engineered ``swap'' channels,
the full Liouvillian is block-additive,

\begin{equation}
\label{E: M replica Liouvillian}
    \mathcal{L}_M = \sum_{i=1}^M \left( \mathcal{S}_i + \sum_{N=1}^{M-i} \mathcal{C}_{i,i+N} \right)
\end{equation}

\paragraph*{Structure and properties.}---Written in matrix form, \(\mathcal{L}\) exhibits a block-diagonal pattern along its
super-operator rows and columns: the diagonal blocks  contain the single-world Lindblad operators, whereas the off-diagonal blocks hold the cross-world channels, each multiplied by the appropriate Boltzmann factor derived
from the generalized KMS condition.  This construction guarantees complete
positivity and trace preservation for every integer $M$, and it naturally
reduces to the familiar two-replica Liouvillian when $M=2$.  Moreover, the explicit decomposition into on-site and hopping pieces provides a transparent
starting point for perturbative or numerical studies of the entropy production as
one approaches the delicate limit $M\to1$.

For a probe environment strongly hybridized to the qubit ($\Gamma_b = \Gamma$), we numerically evaluate the R\'enyi entropies using the hybridization formalism for $M = 2, 3, 4,$ and $5$. In each case, we consider three driving amplitudes—weak, intermediate, and strong—corresponding to $\Omega/\Gamma_e = 0.001, 1, 2$, respectively. Figure (\ref{F: Lambda vs M}) shows the resulting families of curves for various temperatures $\omega/T_e$. For a fixed qubit frequency, larger values of $\omega/T_e$ correspond to colder environments, depicted by bluer curves, whereas smaller values correspond to warmer environments, shown in redder curves.

In the strong‐driving regime, the hotter‐environment curves exhibit a steeper dependence on $M$, causing results for different temperatures to converge, while remaining more sensitive to variations in $M$. In contrast, weak driving cases show more widely separated dependence on $M$ with the colder environment showing a stronger dependence on $M$ and the hotter results being comparatively insensitive to it.

The dependency on $M$ can be useful for estimating the von Neuman intropy. 

\subsection{von Neumann entropy current}
\label{subsec von}

Plotting the hybridized R\'enyi-$M$ entropy flows in Fig.~(\ref{F: Lambda vs M}) can, in principle, be used to extract the von Neumann entropy. However, this would require knowledge of the entropy flows for non-integer $M$ values, which is not directly accessible. Instead, we fit the $M$-dependence of the entropy flows using a functional form that is valid and consistent across all temperatures, with fitting coefficients that depend on temperature and driving amplitude, but not on $M$. Our best approximation is
\begin{equation}
-\Lambda_{0}(M) \approx a \left(M + \frac{b}{M+c} - 1 - \frac{b}{1+c} \right),
\label{eq. fittingfunc}
\end{equation}
where $a$, $b$, and $c$ are fitting parameters determined by $\Omega/\Gamma_b$ and $\omega/T_e$. In Appendix (\ref{app.fitting}) we present the fitting paramters versus temperature for the three cases of weak, intermediate, and strong driving amplitudes.   

The solid lines in Fig.~(\ref{F: Lambda vs M}) represent the fitted curves for each case. This functional form is particularly useful for estimating the flow near $M=1$. In the limit of a single replica, the von Neumann entropy production rate is obtained from
\begin{equation}
S_{vN}=-\partial_{M}\Lambda_{0}\big|_{M \to 1} = a \left(1 - \frac{b}{(1+c)^2}\right).
\end{equation}

\begin{figure}[t]
    \centering
\includegraphics[width=0.49\textwidth]{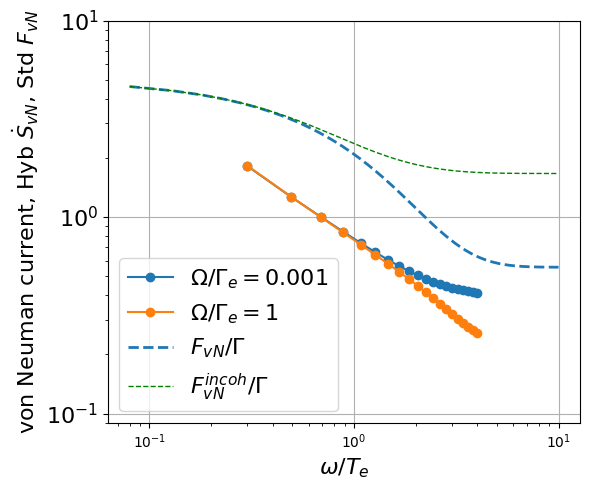} \vspace{-0.3cm}
    \caption{Von Neumann entropy flow derived from strong interaction method for a hybridized system with $\Gamma_b = \Gamma_e$ as a function of $\omega/T_e$, used to evaluate weak (blue dots) and strong (orange dots) driving amplitudes. The dashed line in blue dashed color labeled as $F_{vN}$ represents the weak‐coupling probe estimate from Eq.~\eqref{eq.qubitfvn}. Thinner (green) dashed line labeled with $F_{vN}^{\rm{incoh}}$ denotes the incoherent von Neumann entropy flow without considering the cross-world contributions.}
    \label{fig:svnn}
\end{figure}

Figure~\ref{fig:svnn} shows the von Neumann entropy as a function of $\omega/T_e$ for several driving amplitudes. The corresponding von Neumann entropy estimates obtained from the fitting procedure are also presented in Fig.~\ref{fig:svnn} for weak and strong driving. The dashed line lablled with $F_{vN}$ represents the von Neumann entropy current evaluated directly from Eq.~\eqref{eq.qubitfvn}, which shows quantum coherent contribution in comparison to incoherent-only part of the flow presented by  thinner dashed line labeled as $F_{vN}^{\rm{incoh}}$. 

Although we employed a simple fitting function to estimate the entropy current for non-integer $M$, the resulting von Neumann entropy closely matches the prediction of Eq.~\eqref{eq.qubitfvn}, aside from a slight shift and deformation.  

In all curves, lowering the reservoir temperature suppresses the entropy, and in the case of the new hybridized entropy-flow estimation, this suppression follows an almost power-law behavior, particularly in the strong-driving limit. The primary source of the observed deformation is that the calculation was performed for $\Gamma_b = \Gamma$, which makes the direct evaluation of entropy flow from Eq.~\eqref{eq.qubitfvn} less accurate. The power-law dependence of the hybridized von Neumann entropy can be expressed as
\begin{equation}
S_{vN} \propto \left(\frac{\omega}{T_e}\right)^{\beta},
\end{equation}
with $\beta \approx -1$. This indicates that, in the strong-coupling regime, lowering the temperature leads to a faster suppression of entropy compared to previous calculations based on other formalisms, with the effect being more pronounced at higher driving amplitudes. Similarly one can expect that the increase of external temperature helps to speed up in entropy production.

\section*{Conclusion}

We have introduced a generalized multi-replica Keldysh–Ansari–Nazarov (KAN) formalism capable of quantifying entropy and information flow in quantum–classical hybrid systems for arbitrary coupling strengths. This approach extends beyond conventional weak-coupling treatments by explicitly incorporating strong hybridization, where quantum and classical subsystems share substantial degrees of freedom. By formulating the multi-replica master equation within a tensor-network representation, the method consistently accounts for both intra-replica (same-world) and inter-replica (cross-world) processes, the latter enabling coherence transfer between replicas.

The formalism has been applied to a driven qubit coupled to two reservoirs, serving as a minimal model to contrast weak- and strong-coupling regimes. In the weak-coupling limit, results from the hybridized framework agree with standard expressions; however, in the strong-coupling regime, significant deviations emerge, particularly at elevated drive amplitudes. Numerical analysis demonstrates that hybridization can either enhance or suppress entropy currents, with the effect being most pronounced at low reservoir temperatures where quantum coherence dominates.

The extension to an arbitrary number of replicas $M$ permits the estimation of the von Neumann entropy current by fitting the $M$-dependence of R\'enyi entropy flows. The extracted scaling follows a power-law suppression with temperature, $S_{\mathrm{vN}} \propto (\omega/T_e)^{\beta}$, with $\beta \approx -1$. This reveals a more rapid entropy suppression in the strong-coupling regime than predicted by earlier formalisms, especially under strong driving conditions.

The observed further reduction of entropy flow in the hybridized system reveals a distinctly quantum-mechanical mechanism by which information–energy exchange can be slowed down. In this regime, quantum coherence and hybridization jointly act to inhibit the net transfer of entropy, effectively introducing a ``bottleneck'' in the thermodynamic process. Such a slowdown has direct implications for quantum thermal devices: in a quantum refrigerator or heat engine, a single qubit operating as an entropy valve between two heat reservoirs can significantly delay the irreversible flow of entropy from the hot to the cold reservoir. This delay not only enhances control over thermal gradients but could also improve efficiency by extending the time window in which useful work can be extracted before thermal equilibrium is reached.

Overall, the proposed framework bridges the methodological gap between weak- and strong-coupling descriptions, providing a robust and versatile tool for accurately modeling entropy dynamics in realistically hybridized quantum devices. These results have direct implications for the design and optimization of solid-state quantum processors, quantum thermodynamic machines, and other platforms where quantum–classical hybridization plays a central role.

\section*{Acknowledgement}
The initial idea for this project originated in 2015 in Delft, and its realization has taken nearly a decade. Over this period, numerous talented colleagues have contributed through insightful discussions and, in some cases, direct involvement in parts of the research as coauthors.  M.A. gratefully acknowledges valuable discussions with Lee Smolin, Jukka Pekola, Yasuhiro Utsumi, and Udo Seifert.

\bibliography{references,ref}
\newpage

\appendixpage
\appendix

\section{Generalized Correlators and KMS relation}
\label{Appendix Generalized Correlator}

With the interaction defined, we can introduce the multi-world correlation functions in the probe reservoir

\begin{equation}
\label{E: XX multicorrelator}
    S^{N,M}_{mn,pq} (\tau) = \expval{\!\!\expval{\hat{X}_{mn}(0) \hat{\rho}_A^N \hat{X}_{pq} (\tau) \hat{\rho}_A^{M-N}}\!\!}
\end{equation}

and their half-sided Fourier transform

\begin{equation}
\label{E: Multicorrelator one-sided FT}
    \int_0^\infty d\tau S^{N,M}_{mn,pq} (\pm \tau) e^{i\omega \tau} = \frac{1}{2} S^{N,M}_{mn,pq} (\pm \omega) \pm i\Pi^{N,M}_{mn,pq} (\pm \omega)
\end{equation}

which play a crucial role in the evaluation of the contributing diagrams, especially due to the generalized Kubo-Martin-Schwinger relation \cite{AnsariNazarovEngines}

\begin{equation}
\label{E: Generalized KMS}
    S_{mn,pq}^{N,M} (\omega) = \overline{n} (M\beta_A \omega) \chi_{mn,pq}(\omega) e^{N \beta_A \omega}
\end{equation}

that relates them to the dynamical susceptibilities of the probe reservoir. Here, $\overline{n}$ is the Bose distribution and $\beta_A$ the inverse temperature of the probe.

\[\Delta_C = \Pi_{01,10}^{0,1}(\omega_{k})-\Pi_{10,01}^{0,1}(-\omega_{k}) \]

\begin{widetext}

\section{Liouvillian Operator for $M=2$}
\label{appendix. Liuv}

In this section we express the total Liouvillian matrix associated with the $M=2$ replicas of a two level qubit coupled to two reservoirs. Since density matrix of two-replicated qubits has 16 elements, the Liouvillian matrix carries 16x16 elements, which will be presented below following the Keldysh diagrams we described in Section III and IV. For brevity we use several symbols to make the matrix fit into the paper size. 
\[
A=\left(3+ 0.5e^{2\omega/T}+0.5 e^{\omega/T}\right) \bar{n}, \quad B=\frac{1.5\,e^{2\omega/T}+1.5\,e^{\omega/T}+1}{e^{\omega/T}-1}, \quad C=\frac{e^{2\omega/T}+e^{\omega/T}+2}{e^{\omega/T}-1}
\]
and $\exp(\omega/T)+1=d$.

$x=i \frac{\Omega}{2 \Gamma}$. 
$S_i^{N,M}=\gamma_i e^{N\theta_i}\bar{n}(M \theta_i)$ for $\theta_i=\hbar \omega/k_B T_i$, and $\gamma_b=\Gamma_b/\Gamma$ and $\gamma_e=1$, $S^0=(S_b^{0,2} +  S_e^{0,1})$ and $S^1=(S_b^{1,2} +  S_e^{1,1})$. 

\scriptsize
\[
\left[\begin{array}{cccccccccccccccc}
00,00 & 00,01 & 00,10 & 00,11 & 01,00 & 01,01 & 01,10 & 01,11 & 10,00 & 10,01 & 10,10 & 10,11 & 11,00 & 11,01 & 11,10 & 11,11 
\\
-2 S^0 & ix & ix & 0 & -ix &  S^1 & S_b^{2,2} & 0 & -ix & S_b^{2,2} & S^1 & 0 & 0 & 0 & 0 & 0\\
ix & -A & -S_b^{1,2} & ix & 0 & -ix & 0 & S_b^{2,2} & 0 & -ix & 0 & S^1 & 0 & 0 & 0 & 0\\
ix & -S_b^{1,2} & -A & ix & 0 & 0 & -ix & S^1 & 0 & 0 & -ix & S_b^{2,2} & 0 & 0 & 0 & 0\\
0 & ix & ix & -2 C & 0 & 0 & 0 & -ix & 0 & 0 & 0 & -ix & 0 & 0 & 0 & 0\\
-ix & 0 & 0 & 0 & -A & ix & ix & 0 & -S_b^{1,2} & 0 & 0 & 0 & -ix & S_b^{2,2} & S^1 & 0\\
n_2 & -ix & 0 & 0 & ix & -2 C & -S_b^{1,2} & ix & 0 & -S_b^{1,2} & 0 & 0 & 0 & -ix & 0 & S^1\\
S_b^{0,2} & 0 & -ix & 0 & ix & -S_b^{1,2} & -2 C & ix & 0 & 0 & -S_b^{1,2} & 0 & 0 & 0 & -ix & S_b^{2,2}\\
0 & S_b^{0,2} & n2 & -ix & 0 & ix & ix & -B & 0 & 0 & 0 & -S_b^{1,2} & 0 & 0 & 0 & -ix\\
-ix & 0 & 0 & 0 & -S_b^{1,2} & 0 & 0 & 0 & -A & ix & ix & 0 & -ix & S^1 & S_b^{2,2} & 0\\
S_b^{0,2} & -ix & 0 & 0 & 0 & -S_b^{1,2} & 0 & 0 & ix & -2 C & -S_b^{1,2} & ix & 0 & -ix & 0 & S_b^{2,2}\\
n2 & 0 & -ix & 0 & 0 & 0 & -S_b^{1,2} & 0 & ix & -S_b^{1,2} & -2 C & ix & 0 & 0 & -ix & S^1\\
0 & n2 & S_b^{0,2} & -ix & 0 & 0 & 0 & -S_b^{1,2} & 0 & ix & ix & -B & 0 & 0 & 0 & -ix\\
0 & 0 & 0 & 0 & -ix & 0 & 0 & 0 & -ix & 0 & 0 & 0 & -2 C & ix & ix & 0\\
0 & 0 & 0 & 0 & S_b^{0,2} & -ix & 0 & 0 & n2 & -ix & 0 & 0 & ix & -B & -S_b^{1,2} & ix\\
0 & 0 & 0 & 0 & n2 & 0 & -ix & 0 & S_b^{0,2} & 0 & -ix & 0 & ix & -S_b^{1,2} & -B & ix\\
0 & 0 & 0 & 0 & 0 & n2 & S_b^{0,2} & -ix & 0 & S_b^{0,2} & n2 & -ix & 0 & ix & ix & -2 S_b^{2,2}-2 S_e^{1,1}

\end{array}\right]
\]

The eigenvalues of the Liouvillian matrix can be computed and are shown in Fig.~\ref{fig:eigen}. Panels (a) and (b) display the real parts, while panels (c) and (d) show the imaginary parts. These results are evaluated for various drive amplitudes $\Omega/\Gamma$, ranging from weak to strong driving, at a cold temperature that is still warmer than the probe temperature. The left panels (a, c) correspond to a hybridized system with weak coupling, $\Gamma_b = \Gamma/20$, whereas the right panels (b, d) correspond to the strong‐coupling case, $\Gamma_b = \Gamma$. Notably, the Liouvillian spectrum exhibits several exceptional points, where two eigenvalues become degenerate at specific driving amplitudes.

\begin{figure}
    \centering
    \includegraphics[width=0.49\linewidth]{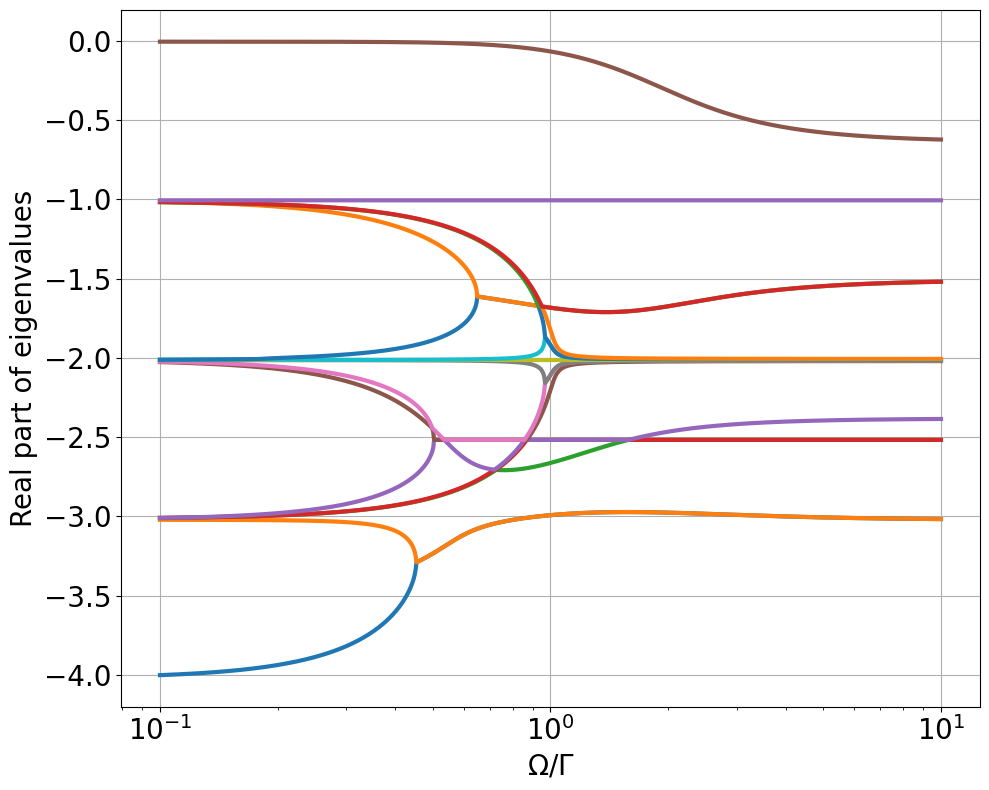}
    \includegraphics[width=0.49\linewidth]{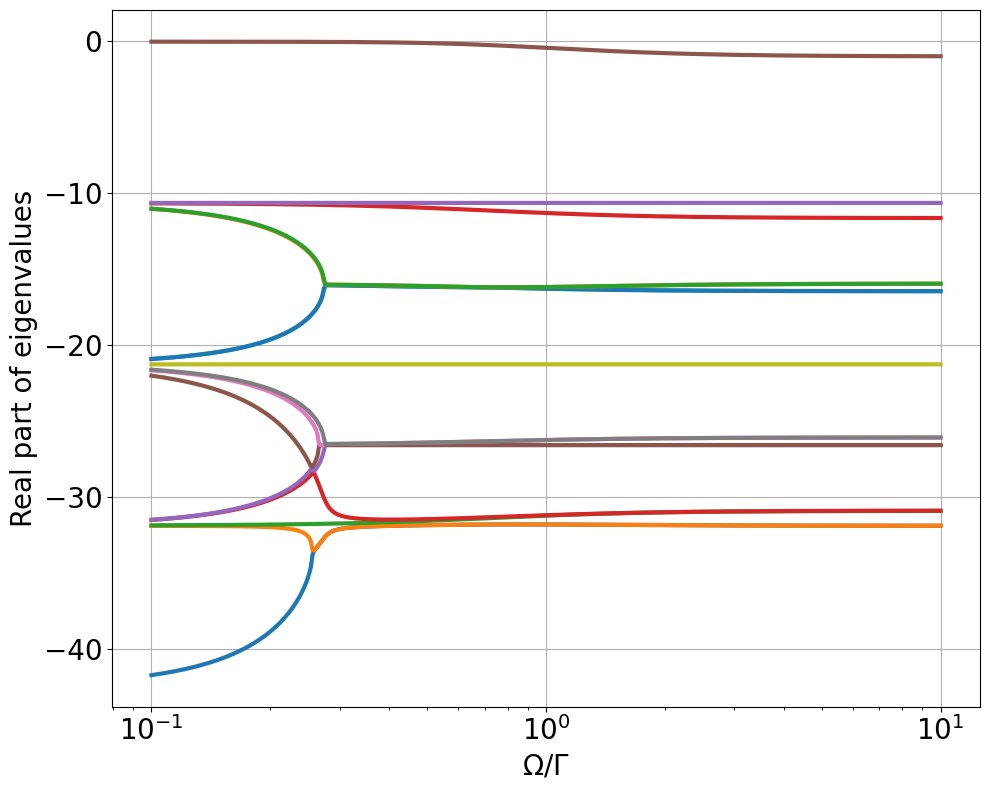}\\
    \includegraphics[width=0.49\linewidth]{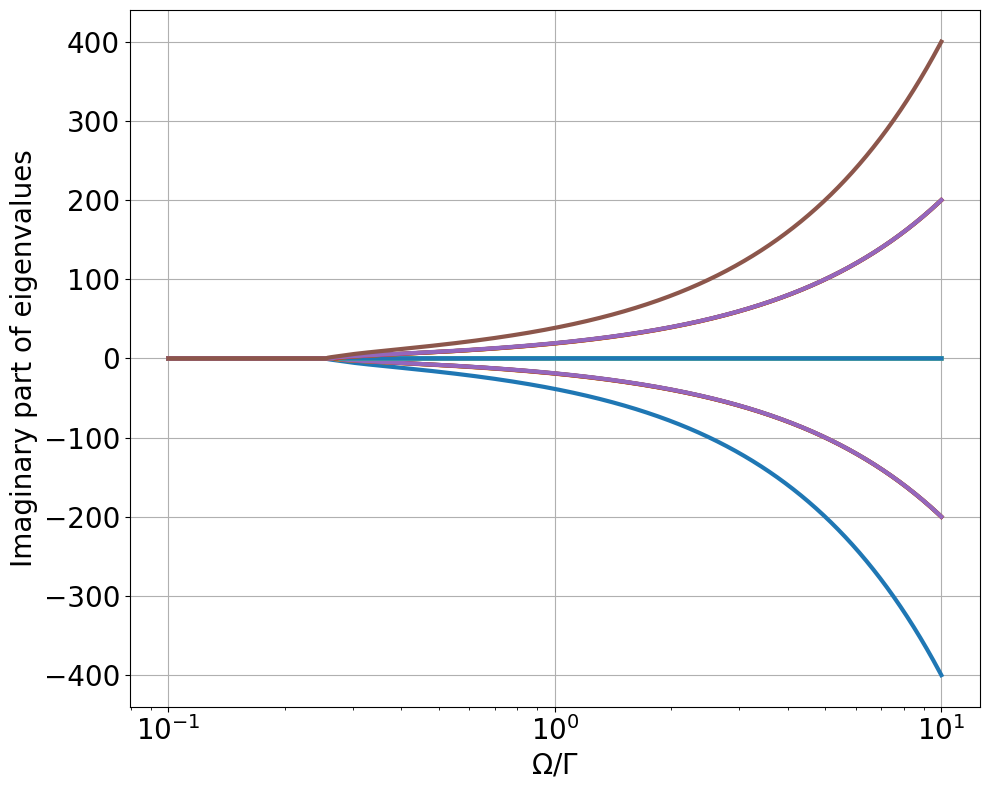}
    \includegraphics[width=0.49\linewidth]{Imalleigenweak5.png}
    \caption{Enter Caption}
    \label{fig:eigen}
\end{figure}

\end{widetext}

\section{Fitting Parameters}
\label{app.fitting}

\begin{figure*}
    \centering
    \includegraphics[width=0.55\linewidth]{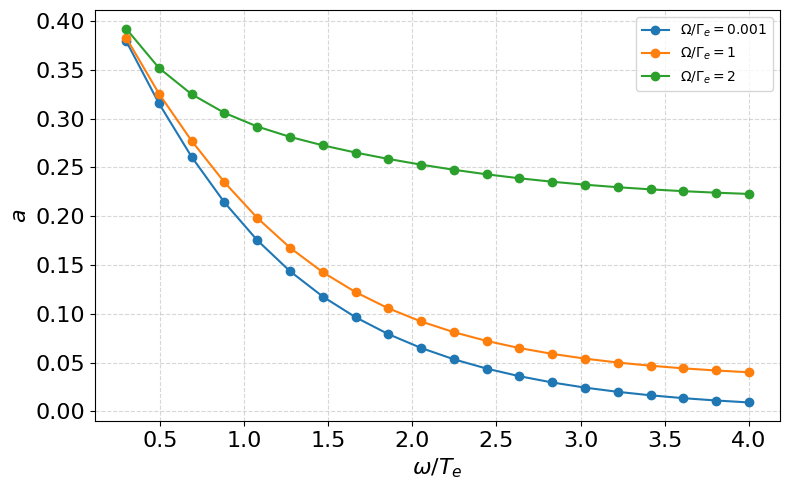}\\
    \includegraphics[width=0.55\linewidth]{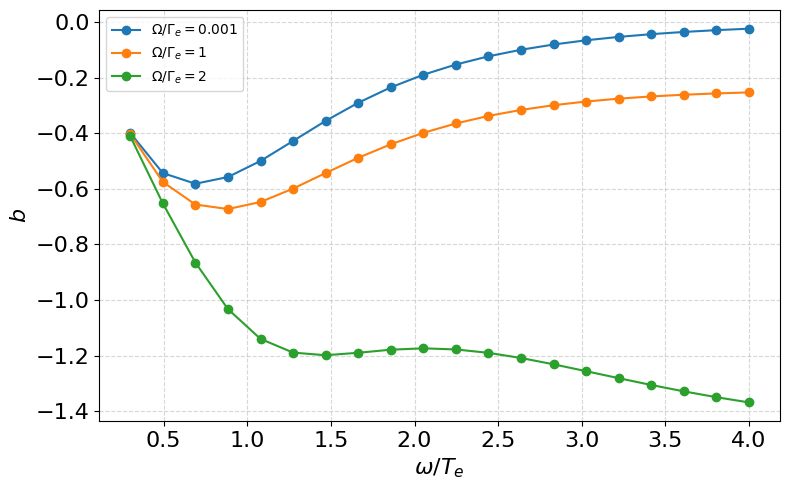}\\
    \includegraphics[width=0.55\linewidth]{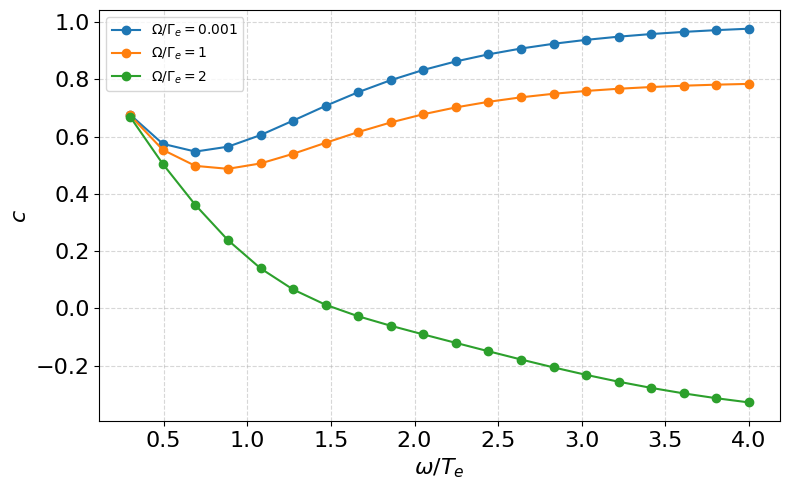}
    \caption{Fitting parameters of Eq. (\ref{eq. fittingfunc}) for different temeprature and driving amplitudes.}
    \label{fig:abcff}
\end{figure*}

Evaluation of R\'enyi entropy current of $M$ replicas versus $M$  from strong coupling regime where the probe reservoir is coupled with the same coupling strength as the real encironment to the quantum system, i.e. $\Gamma_b=\Gamma_e$. Fitting function (\ref{eq. fittingfunc}) fills the flow for interpolating points associated with non-integer $M$. In Fig. (\ref{fig:abcff})  we consider the weak, intermediate, and strong driving amplitudes associated with  $\Omega/Gamma=0.001, 1, 2$ 
 and show how the fitting parameters $a,\ b,$ and $c$ depend on the qubit frequency divided by the environment tempearture $\omega/T_e$.

\end{document}